\documentclass[12pt]{article}
\usepackage{graphicx}    
\usepackage{multicol}   
\usepackage{amsmath}
\usepackage{latexsym}
\usepackage{float}
\usepackage{amssymb}
\usepackage[latin1]{inputenc}
\newcommand{\req}[1]{Eq.(\ref{#1})}
\newcommand{\e}{\mbox{e}}
\renewcommand{\d}{\mbox{d}}

\newcommand{\ol}[1]{\overline{#1}}
\newcommand{\sfrac}[2]{\mbox{$\frac{#1}{#2}$}}
\newcommand{\av}[1]{\left\langle{#1}\right\rangle}
\renewcommand{\phi}{\varphi}
\renewcommand{\epsilon}{\varepsilon}

\begin{document}
\bibliographystyle{unsrt}

\noindent
{\Large\bf Glassy Dynamics and Aging in Disordered \\[2mm] Systems}\\[3mm]
{Heinz Horner}\\[3mm]
{Institut f\"ur Theoretische Physik, Ruprecht-Karls-Universit\"at Heidelberg}\\
{Philosophenweg 19,  D-69120 Heidelberg, Germany}\\
{\texttt{horner@tphys.uni-heidelberg.de}}\\[3mm]
\noindent WE-HERAEUS-FERIENKURS F\"UR PHYSIK 

\noindent Collective Dynamics in Nonlinear and Disordered Systems

\noindent  Chemnitz, Aug. 26. - Sept. 6. 2002 

\setcounter{page}{0}
\tableofcontents 
\newpage

%
\noindent
{\Large\bf Glassy Dynamics and Aging in Disordered \\[2mm] Systems}\\[7mm]
{Heinz Horner}\\[3mm]
{Institut f\"ur Theoretische Physik, Ruprecht-Karls-Universit\"at Heidelberg}\\
{Philosophenweg 19,  D-69120 Heidelberg, Germany}\\
{\texttt{horner@tphys.uni-heidelberg.de}}


\section{Introduction}
A great number of disordered systems exhibit very long relaxation times as some critical
temperature is approached. Below this temperature equilibrium is no longer reached  within
finite time and  the behaviour of such systems becomes non ergodic. Among those are diluted alloys
of magnetic ions in a non magnetic matrix, so called \index{spin-glass}spin-glasses
\cite{SG-rev}, 
\index{supercooled liquid}supercooled liquids \cite{Glas-rev} entering some glassy state at low
temperature or particles moving in a random potential exhibiting a transition from diffusion to
\index{creep}creep or
\index{pinning}pinning \cite{creep,Ho96}. The temporal behaviour of such systems is often
referred to as
\index{glassy dynamics}glassy dynamics. Below the characteristic temperature \index{aging}aging is
observed \cite{aging}. If, for instance, a spin-glass is cooled in a magnetic field and kept for
some waiting time, the complete decay of the magnetisation after the field has been switched off,
is hindered for times of the order of the \index{waiting time}waiting time. Similar phenomena are
observed for the deformation of glasses under the influence of applied forces.

Glassy dynamics can also be of relevance for systems and phenomena outside physics. For some
\index{combinatorial optimisation}combinatorial optimisation problems 
\index{simulated annealing}simulated annealing is used in order to find good solutions. This means
that some stochastic dynamics, characterized by some ``temperature'', is used in order to find
minima of a properly defined cost function. If such a system exhibits a freezing transition, it
is not effective to spend much time in simulated annealing below this temperature and
investigating the dynamics slightly above the freezing transition is much more efficient
\cite{Ho92}. Other examples of systems of interest outside physics are models of 
\index{neural network}neural networks or certain aspects of markets and other economical
systems\index{econophysics} \cite{Minor}. 

The systems mentioned above are classical. There is also interest in quantum mechanical
disordered systems, for instance in certain spin-glasses or in interacting tunnelling systems
being of relevance for glasses at very low temperatures. This lecture deals, however, with
classical dynamical systems only. 

A breakthrough of our understanding of the low temperature properties of glassy systems came
from \index{replica theory}replica theory and the \index{replica symmetry breaking}replica
symmetry breaking scheme proposed by Parisi \cite{Parisi-rev}. This theory focuses on the
evaluation of the free energy averaged over some frozen disorder. The picture emerging from this
theory is a decomposition of phase space into so called pure states or \index{ergodic
components}ergodic components, separated by barriers of infinite hight in the thermodynamic limit.
Investigating the overlap among those states, an
ultrametric\index{ultrametricity} organization reveals.

An alternative treatment of systems with quenched disorder is via
dyn\-amics \cite{SZ82,Ho84,Bou98,Cug03}. This will be the subject of this lecture. As it turns out
there are many features and results common to both treatments, but there are also differences
which will be discussed.  

In the first part I present some of the systems of interest, in particular spin-glasses,
supercooled liquids and glasses, drift, creep and pinning of a particle in a random
potential, neural networks, graph partitioning as an example of combinatorial optimisation,
the K-sat problem, a prototype problem from computer science, and finally the minority game as
a model for the behaviour of agents on markets.

The second part deals with the dynamics of the spherical \index{p-spin-glass}$p$-spin-glass
with long ranged interactions. This model is a prototype for glassy dynamics. The equations
of motion for correlation- and response-functions are derived and equilibrium solutions above the
\index{freezing temperature}freezing temperature are investigated. Below the freezing temperature
the system behaves for short time as if it were in equilibrium, for longer time its off
equilibrium nature becomes apparent and correlation- and response-functions depend on waiting
time. This is the manifestation of aging. 

At the end of this part I discuss the so called \index{crossover region}crossover region and
aging in glasses. Within \index{mode coupling theory}mode coupling theory, which is discussed in
the contribution of R. Schilling \cite{Sch03} in this volume, and which shows great similarity to
the $p$-spin model, a dynamical transition is found at some critical temperature $T_c$. This is
sometimes referred to as ``ideal glass transition''. Comparison with experiments shows that this
transition is actually smeared out and the fitted
$T_c$ is well above the actual glass transition region. Some smearing out also results from
an extended version of the mode coupling theory, which applies, however, to equilibrium
only. I am going to discus a modified version of the $p$-spin model, which is able to take
account for this rounding as well as aging. \\

\section{Examples of Disordered Systems}

\subsection{Spin-glasses}

\begin{figure}[ht]
\centering
\includegraphics[width=6cm]{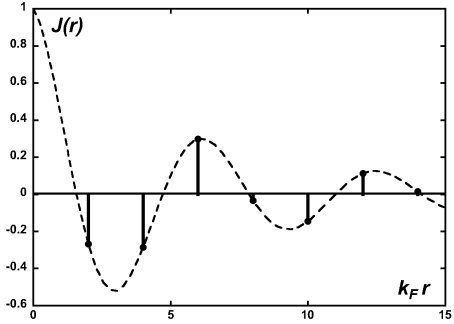}
\caption{RKKY-interaction }
\label{RKKY} 
\end{figure}

Typical \index{spin-glass}spin-glasses are diluted alloys of magnetic ions in a non magnetic
matrix, e.g.
$Ag_{1-x}Mn_x$ with $x\sim3\%$. The magnetic ions interact via the RKKY-interaction which is
mediated through the conduction electrons. It is of the form
\begin{equation}
J(r)\sim r^{-3} \cos(k_f r)
\end{equation}
shown in Fig.\ref{RKKY}.
For random distances $r_{ij}$ between pairs of magnetic ions, their exchange interactions
$J_{ij}\!=\!J(r_{ij})$ are also random variables including positive and negative values.
This leads to \index{frustration}frustration, as shown in the insert of Fig.\ref{RKKY} for a
triple $i\,j\,k$ of ions with $J_{ij}>0$, $J_{i,k}>0$ and $J_{j,k}<0$.

\begin{figure}[ht]
\centering
\includegraphics[width=6cm]{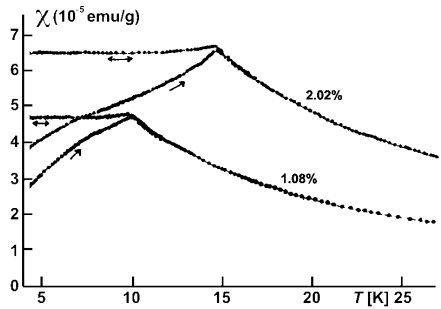}
\caption{Susceptibility of $\underline{\rm Cu}{\rm Mn_x}$ \cite{Nag79} for different values of
the concentration $x$ of the magnetic ions.}
\label{X-FC-ZFC} 
\end{figure}

\begin{figure}[ht]
\centering
\includegraphics[width=6cm]{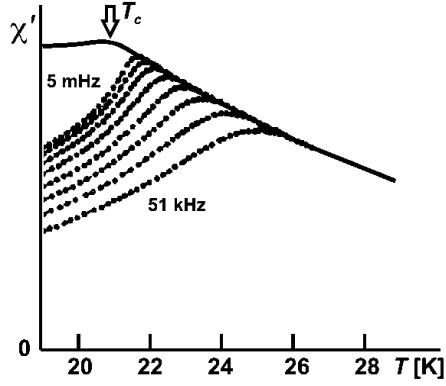}
\caption{Real part of the susceptibility $\chi'(\omega)$ for ${\rm
Fe_{0.5}MN_{0.5}TiO_3}$ \cite{Gun91}. The critical temperature is $T_g=20.7\,{\rm[K]}$}
\label{SG-X(w)} 
\end{figure}

At high temperatures spin-glasses typically show a Curie-law for the magnetic
susceptibility 
$\chi\sim 1/T$, as shown in Fig.\ref{X-FC-ZFC} for $\underline{\rm Cu}{\rm Mn_x}$. Below
some freezing temperature $T_g$ the system is no longer in equilibrium. If the system is
cooled in a small field, the magnetisation $M=\chi_{FC}B$ stays more or less
constant ($a$ and $c$ in Fig.\ref{X-FC-ZFC}). The ratio $\chi_{FC}=M/B$ is called
\index{field cooled susceptibility}field cooled susceptibility. 
\\
If, however, the system is cooled without applied field, and
the field is applied only after some waiting time $t_w$, the resulting magnetisation is
reduced, i.e. the \index{zero field cooled susceptibility}zero field cooled susceptibility
$\chi_{ZFC}=M/B<\chi_{FC}$ ($b$ and
$d$ in  Fig.\ref{X-FC-ZFC}). Approaching the critical temperature $T_g$ critical slowing down can
be observed, see Fig.\ref{SG-X(w)}.

\begin{figure}[ht]
\centering
\includegraphics[width=7.5cm]{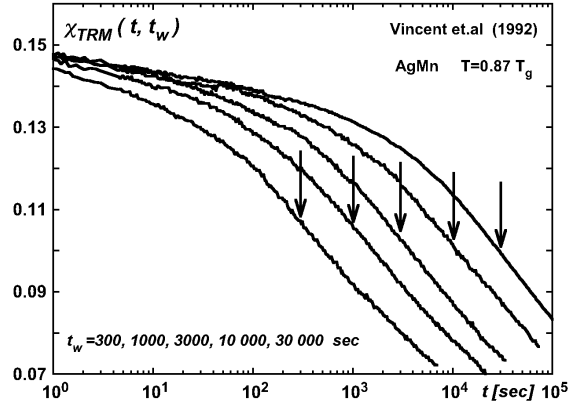} 
\caption{Decay of the remnant magnetisation for various waiting times $t_w$ indicated by
arrows \cite{Ref87}.}
\label{SG-aging} 
\end{figure}

The off equilibrium character of the low temperature phase is most clearly demonstrated
by the decay of the remnant magnetisation. In this experiment the sample is rapidly
cooled in a magnetic field to a temperature $T<T_g$. After some waiting time $t_w$ the
field is switched off, and the decay of the magnetisation is observed as function of time
$t$ elapsed after removal of the field. Following some rapid initial decay, not shown in
Fig.\ref{SG-aging}, the magnetisation stays almost constant up to time $t\sim t_w$. 

Investigating more elaborate temperature programs $T(t)$ a variety of other striking
memory and \index{aging}aging effects are observed \cite{Vin99}. Similar aging phenomena are also
found in glasses \cite{aging}. 

Theoretical investigations are typically based on \index{Ising model}Ising models with random
interactions. Their energy is\\
\begin{equation}\label{kgby}
H=-\sfrac12 \sum_{i,j} J_{i,j} \sigma_i\sigma_j \qquad\quad \sigma_i=\pm1.
\end{equation}
\\
The interactions $J_{ij}$ are assumed to be gaussian distributed random variables with\\
\begin{equation}
\ol{J_{i,j}}=0 \qquad\quad \ol{J_{i,j}J_{k,l}}
=\sfrac12\{\delta_{i,k}\delta_{j,l}+\delta_{i,l}\delta_{j,k}\}W_{i,j}.
\end{equation}
\\
Calculating e.g. the free energy
\\
\begin{equation}
F=-k_B T\, \ol{\ln\Big(\sum_{\{\sigma=\pm1\}}\e^{-\beta H}\Big)}^J
\end{equation}
the problem arises, that the logarithm of the partition function has to be averaged over the
gaussian disorder. Similar problems exist in evaluating expectation values, e.g.
\\
\begin{equation}
\av{\sigma_i\sigma_j}=\ol{\Big(\sum_{\{\sigma\}}\sigma_i\sigma_j\e^{-\beta H}\Big)\Big/
\Big(\sum_{\{\sigma\}}\e^{-\beta H}\Big)}^J.
\end{equation}
\\
This can be done using the replica trick \cite{Parisi-rev}.

As an alternative, one can examine dynamics in the form of stochastic processes having the
Boltzmann distribution $\sim\e^{-\beta H}$ as stationary solution. For an Ising model,
\index{Glauber dynamics}Glauber dynamics can be used. It is is given by a single spin-flip
master-equation such that
\begin{equation}\label{mbts}
\frac{\d}{\d t}\av{\sigma_i(t)}=\av{\tanh\Big(\beta\sum_i J_{i,j}\sigma_j(t)\Big)}.
\end{equation}
Using a path integral representation of this process, the average over the stochastic interactions
can easily be performed. This will be subject of the second part of this lecture.

\subsection{Supercooled Liquids and Glasses}

\begin{figure}[ht]
\centering
\includegraphics[width=6cm]{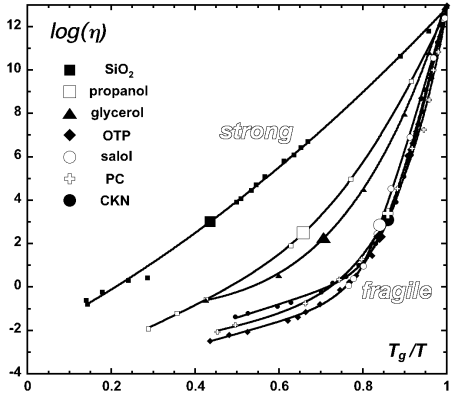}
\caption{Angell plot: Logarithm of the viscosity versus inverse temperature \cite{Ang85}. A
straight line corresponds to Arrhenius behaviour. The fat symbols indicate the MCT-crossover
temperature $T_c$ \cite{MCT}. }
\label{Angell} 
\end{figure}

If a liquid is cooled sufficiently slow, it may avoid crystallisation and enter the state of a
\index{supercooled liquid}supercooled liquid. With decreasing temperature the shear viscosity
$\eta(T)$ increases and reaches a value of $\eta(T_g)\approx 10^{12}\,{\rm [Pa\,sec]}$ defining
the glass temperature
$T_g$. At this value, plastic flow can hardly be observed in laboratory experiments. In some
glasses, so called strong glasses, the viscosity follows more or less an Arrhenius law
$\eta(T)\sim \e^{\beta E_a}$ \cite{Ang85}. A typical strong glass formers is ${\rm SiO_2}$ with an
activation energy  $E_a\approx 4\,{\rm eV}$ corresponding to the binding energy of a covalent
bond. Fragile glass formers, on the other hand, show pronounced deviations from this law. Typical
examples are organic molecular glasses, ionic glasses, polymers or proteins. Some examples are
shown in Fig.\ref{Angell}.

There are other attempts to define a \index{glass transition}glass transition temperature, e.g.
identifying a rapid drop in the specific heat or thermal expansion \cite{Dont}. Actually the
glass temperature defined in one way or another usually depends on the cooling rate, and it is
not clear at all, if some finite
$T_g$ exists in real glasses. There is no need to discuss this further in this lecture, since
theories of the structural glass transitions are covered in the contribution of R.
Schilling \cite{Sch03} in this volume. 

There is, however, one aspect of interest in the present context. This is the ''Ideal Glass
Transition'' found in mode coupling theory (MCT) \cite{MCT,Sch03}. The resulting equations
resemble those found for the dynamics of spherical $p$-spin interaction
spin-glass\index{p-spin-glass} \cite{Kir87,CHS93} to be discussed later in this lecture. This
theory yields diverging time scales as some critical temperature $T_c$ is approached, and non
ergodic behaviour for $T<T_c$. 

\begin{figure}[ht]
\centering
\includegraphics[width=5.6cm]{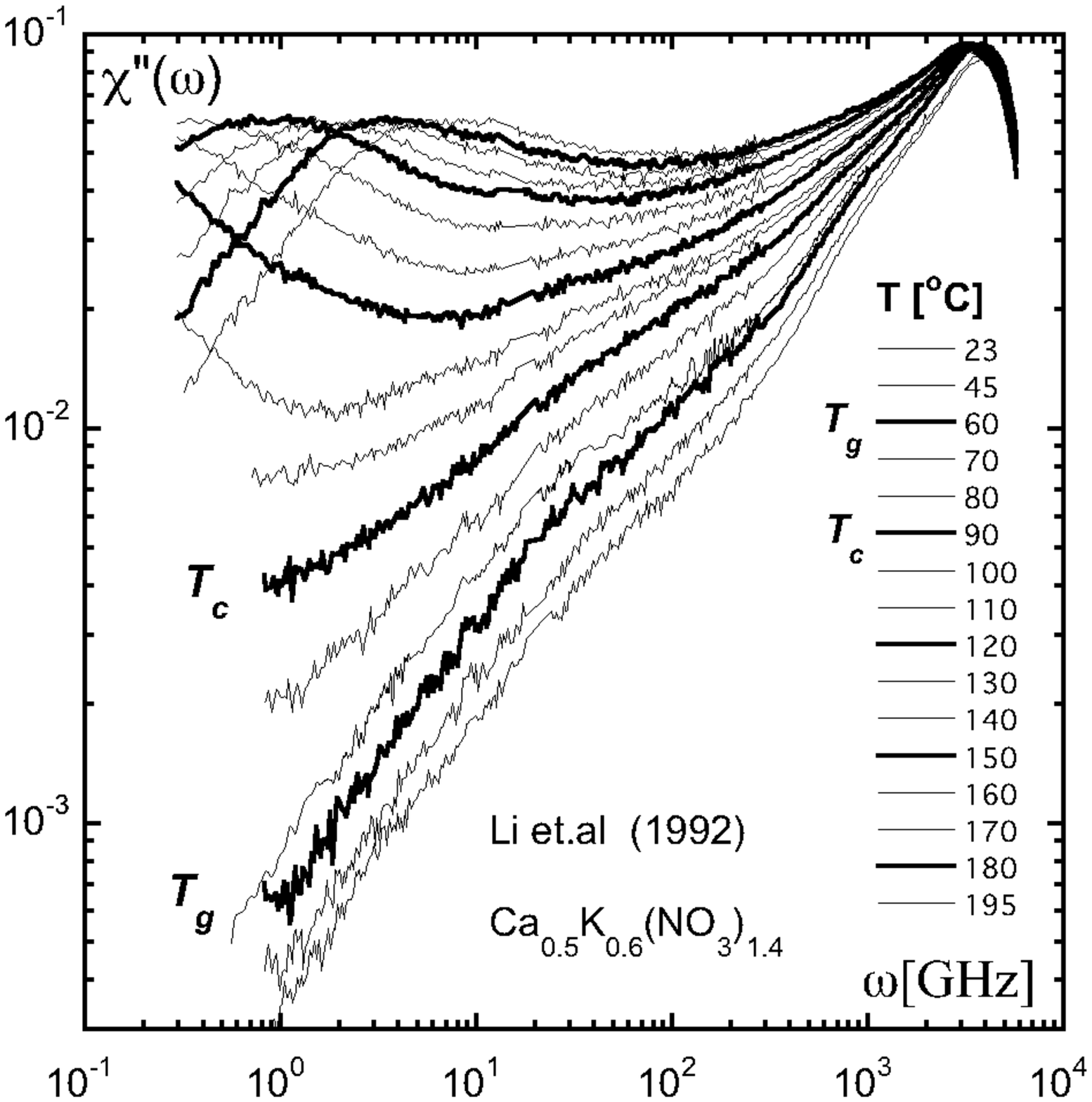}
\includegraphics[width=5.9cm]{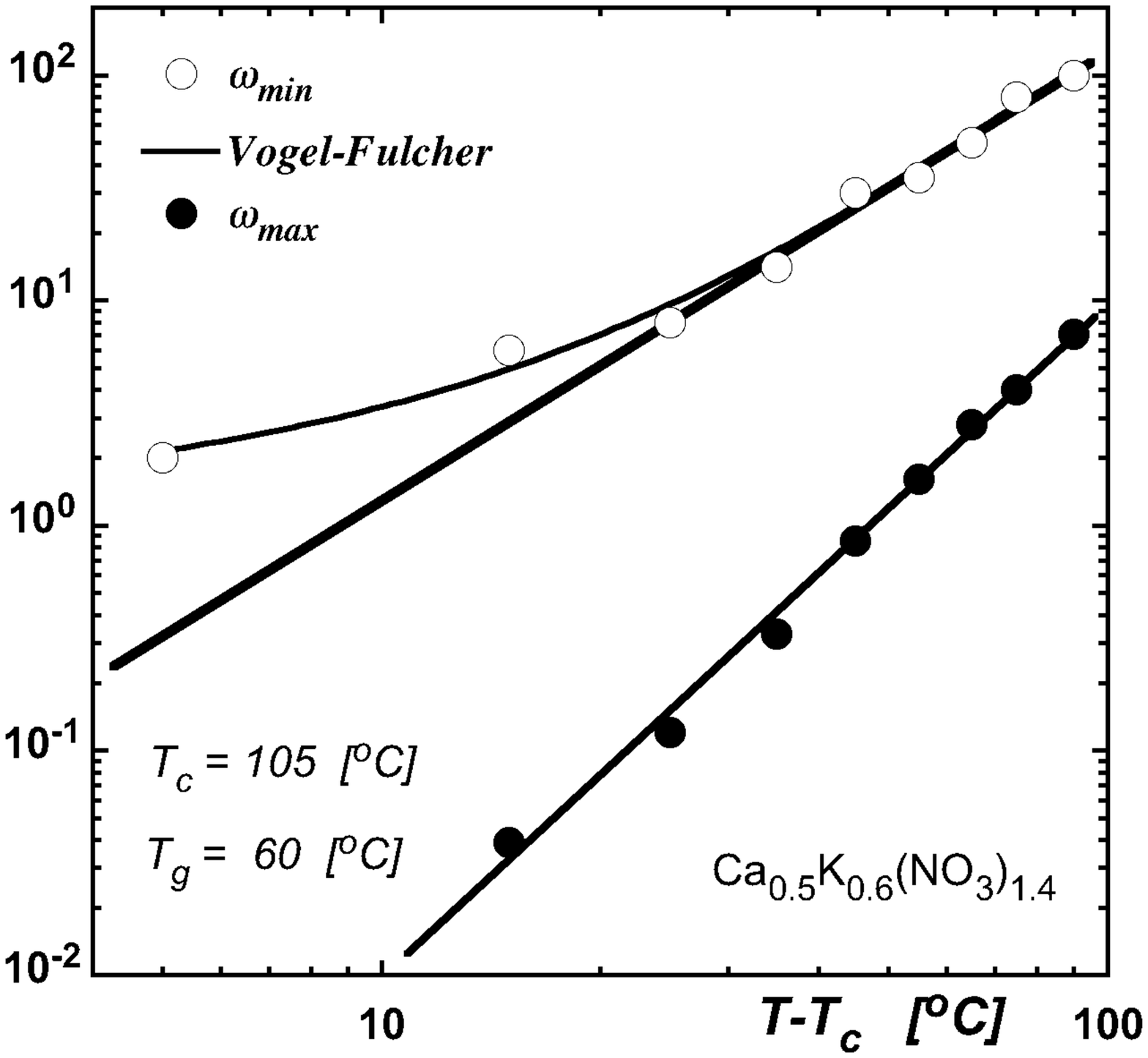}\\
{\bf (a) \hspace{5cm} \ \ \ (b)}
\caption{{\bf (a)} Imaginary part of the dielectric susceptibility $\chi''(\omega)$ of the ionic
glass CKN \cite{CKN92}. The $\alpha$-relaxation shows up as the broad peak at the lower
frequency.  
{\bf (b)} The frequencies of the maximum $\omega_{\rm max}$ and minimum $\omega_{\rm min}$
of $\chi''(\omega)$ fitted to \  $\omega_{\rm max}\sim(T-T_c)^\eta$ \ and \ 
$\omega_{\rm min}\sim(T-T_c)^{\eta'}$.}
\label{CKN} 
\end{figure}

An onset of critical slowing down, usually referred to as $\alpha$-relaxation, can be
observed in fragile glasses, e.g. in the dielectric \index{susceptibility}susceptibility
$\chi''(\omega)$ of the ionic glass CKN \cite{CKN92} shown in Fig.\ref{CKN}.a. Fitting the
critical temperature $T_c$ and other parameters of mode coupling theory, as indicated in
Fig.\ref{CKN}.b, yields the critical temperatures shown in Fig.\ref{Angell}.

\begin{figure}[ht]
\centering
\includegraphics[width=7.5cm]{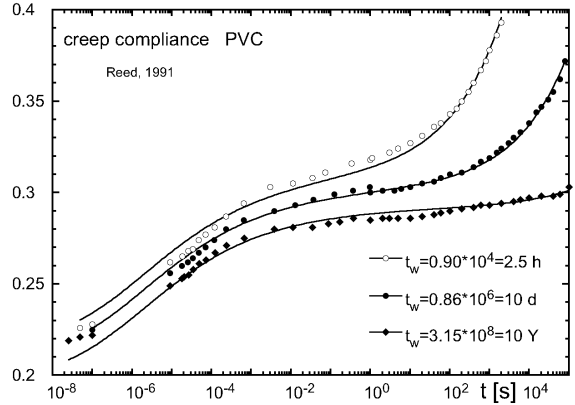}
\caption{Aging observed in PVC \cite{Ree91}. A shear stress is applied after some waiting time
$t_w$ following a quench from the liquid state. Then the resulting deformation is observed as
function of time $t$.}
\label{Reed} 
\end{figure}

Actually an extended version of the \index{mode coupling theory}mode coupling theory
\cite{MCT,Goe99}  yields a rounding of the singularity near $T_c$ and equilibrium behaviour for
temperatures below $T_c$, in contrast to the findings of the simplified mode coupling theory or
the $p$-spin-glass. On the other hand,
\index{aging}aging can be observed in glasses as well, indicating off equilibrium properties at
low temperatures. An example is shown in Fig.\ref{Reed}. We shall come back to this point later.

Assuming pair interactions, the (potential) energy may be written as
\begin{equation}
H=\sfrac12 \!\int\!\d\vec{r}\d\vec{r'}\, V(\vec{r}-\vec{r'})\,n(\vec{r})\,n(\vec{r'})
\end{equation}
where $n(\vec{r})$ is the density at point $\vec{r}$. 

If one is interested in slow dynamics only,
it is sufficient to investigate the limit of overdamped motion, i.e. the Langevin equation
\begin{equation}
\frac{\partial}{\partial t}
n(\vec{r},t)=\vec{\nabla}n(\vec{r},t)\cdot\vec{\nabla}\!\int\!\d\vec{r'}\,
V(\vec{r}-\vec{r'})\,n(\vec{r'},t)
+\vec{\eta}(\vec{r},t)\cdot\vec{\nabla}n(\vec{r},t)
\end{equation}
with fluctuating forces such that
\begin{equation}\label{nsya}
\av{\vec{\eta}(\vec{r},t)}=0 \qquad\quad
\av{\vec{\eta}(\vec{r},t)\cdot\vec{\eta}(\vec{r}',t')}=2\,T\,\delta(\vec{r-r}')\,\delta(t-t').
\end{equation}
Investigating time dependent correlation-functions, the nonlinear term in the above equation is
treated via mode coupling theory. As mentioned, the resulting equations of motions are identical
to those obtained for the $p$-spin-glass \index{p-spin-glass} (in equilibrium). This is
remarkable, because the Hamiltonian of the spin-glass has built in quenched disorder, whereas the
Hamiltonian of the glass does not, and the disorder appears to be self generated.

The relevance of the findings of mode coupling for a glass transition at $T_g$ is not completely
obvious. Other concepts have been proposed to understand the glass transition, e.g.
activated transitions among ``inherent states'' \cite{Stil82} or models with kinetically
constraint dynamics \cite{Jaec91,Ber03}.

%
\subsection{Motion of a particle (manifold) in a random potential}
Another example for glassy dynamics is the motion of a particle\index{drift of a particle} in a
correlated random potential \cite{creep,Ho96}. Using again the limit of strong damping, a Langevin
equation may be used
\begin{equation}
\vec{v}(t)=\frac{\d}{\d t}\vec{r}(t)=-\vec{\nabla}V(\vec{r}(t))+\vec{K}+\vec{\eta}(t)
\end{equation}
where $V(\vec{r})$ is assumed to be a gaussian random potential with moments
\begin{equation}
\ol{V(\vec{r})}=0 \qquad
\ol{V(\vec{r})\,V(\vec{r}')}=\frac{V_o}{(|\vec{r}-\vec{r}'|+a)^\gamma}\, ,
\end{equation}
and $\vec{K}$ is some external force pulling the particle. The thermal noise is again given by
\req{nsya}.
\begin{figure}[ht]
\centering
\includegraphics[width=4cm]{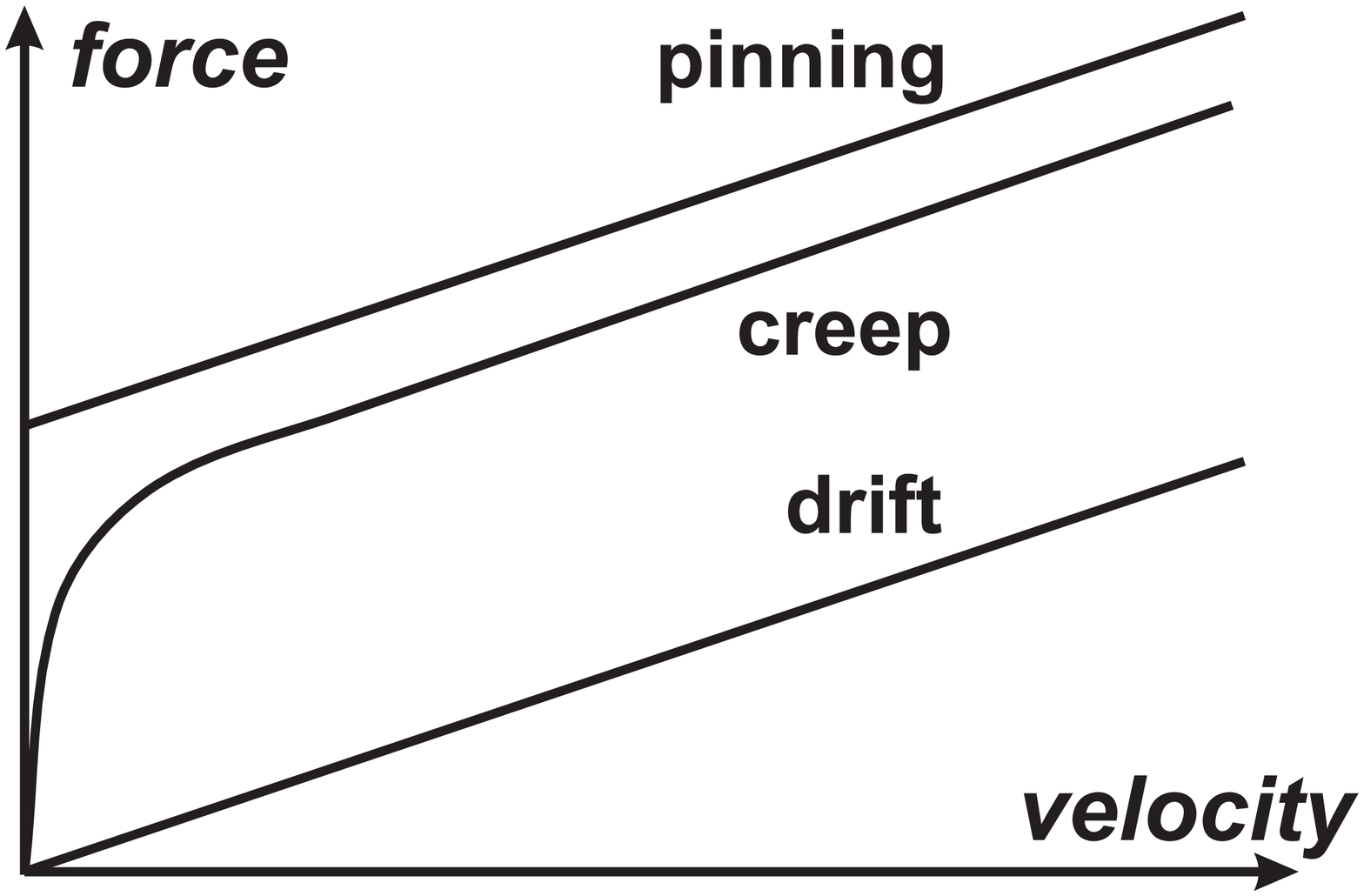}
\caption{Drift, \index{creep}creep and \index{pinning}pinning for a particle moving in a random
potential.}
\label{creep} 
\end{figure}

Depending on temperature $T$ and the
exponent $\gamma$, characterising the range of the random potential, various
types of motion are found \cite{Ho96}, see Fig.\ref{creep}. For $T>T_c(\gamma)$ the average
velocity $v(K)\sim K$, i.e. there is a finite friction constant. For $T<T_c(\gamma)$ and
$\gamma<\gamma_c$ the particle moves only if the force exceeds some critical value $K>K_{\rm
pinning}(T,\gamma)$. For
$T<T_c(\gamma)$ and $\gamma>\gamma_c$ creep is found, i.e. $v(K)\sim K^\eta$ with $\eta<1$.

A similar freezing transition shows up for a chain or some other manifold moving in a random
potential \cite{KiHo93}.

\subsection{Neural Networks}

Considerable effort has been put into the study of formal \index{neural network}neural networks
dealing with random patterns \cite{NNet}, e.g. the Hopfield model \cite{Hop82} as a prototype of
an associative memory. 

Representing the ``on'' and ``off'' state of a neuron by the two states of an Ising spin
$\sigma_i=\pm 1$, the energy is assumed to be that of an Ising spin-glass \req{kgby}. The
couplings $J_{i,j}$ are, however, determined by the patterns to be memorised. 
Let $\xi_i^\mu=\pm 1$ be a set of random patterns, where $i=1\cdots N$ labels the neurons and
$\mu=1\cdots A$ the patterns. Then the choice of the couplings $J_{i,j}$ such that
\begin{equation}
J_{i,j}=\sfrac{1}{N}\sum_\mu \xi_i^\mu\xi_j^\mu
\end{equation}
ensures, that a fixed point of the dynamics, defined by \req{mbts}, exists near each pattern $\mu$
at least as long as $A<\alpha_c N$ with $\alpha_c\approx 0.13$. Each fixed point is surrounded by
some basin of attraction, i.e. if the initial state contains a sufficient part of a pattern, the
complete pattern is reconstructed with a small number of errors. The size of the basin of
attraction depends on the loading $\alpha=A/N$ \cite{Ho89}. For $\alpha>\alpha_c$ the memory is
overloaded and looses its function completely. The dynamics becomes glassy, which is not too
surprising, since the couplings $J_{i,j}$ can be positive as well as negative, resulting in
frustration.

Methods developed for spin-glasses, replica theory and dynamics, have also been used to
investigate this and other types of neural networks. In particular bounds of learning
 \cite{NNet,Ho92} have been determined.

\subsection{Combinatorial Optimisation Problems}

Combinatorial optimisation problems \index{combinatorial optimisation} show up in various
technical applications. The problem is, to find the ``best'' out of a discrete set of states.
Finding the ground state of an Ising spin-glass or solving the travelling sales man
problem, i.e. finding the shortest path connecting a set of towns, are examples. In many cases the
problem is $NP$-complete, i.e. the optimal solution can not be found with computational effort
growing with the size $N$ of the problem to some power. This means that for larger systems it is
virtually impossible to find the optimal solution. In many applications it might, however, be
sufficient to find good solutions in polynomial time. One of the standard procedures is 
\index{simulated annealing}simulated annealing \cite{Kir83}. It is implemented by constructing a
cost function (energy), and using some stochastic dynamics, characterised by a temperature. This
temperature is slowly decreased, and hopefully states with low values of the cost function are
found. 

Actually such a system might undergo a transition to glassy dynamics below some critical
temperature $T_c$. If this is the case, extending simulated annealing to $T<T_c$ is not
efficient, and searching for good solutions at $T\approx T_c$ is more effective \cite{Ho92}.

A standard problem in this context is \index{graph bipartitioning}graph bipartitioning
\cite{Fu86}. Assume electronic components $i=1\cdots N$ have  to be placed on two chips $A$ and
$B$. The aim is to place them such that the number of connections between $A$ and $B$ is minimal.
A cost function for this problem can be defined by mapping it to an Ising system with
\begin{equation}
\sigma_i=1 \quad {\rm if \;\; }i{\rm \;\;on \;\;}A \qquad
\sigma_i=-1 \quad {\rm if \;\; }i{\rm \;\;on \;\;}B.
\end{equation} 
and interactions
\begin{equation}
J_{ij}=1 \quad {\rm if \;\;} i {\rm \;\; and \;\;}j{\rm\;\; are\,connected}\qquad
J_{i,j}=0 \quad {\rm else}.
\end{equation}
The resulting cost function is
\begin{equation}
H=-\sfrac12\sum_{i,j}J_{i,j}\,\sigma_i\,\sigma_j .
\end{equation}
For the process of simulated annealing again Glauber dynamics, \req{mbts}, may be
used.

\subsection{Random $K$-sat Problem}

Another standard $NP$-problem is the \index{K-sat problem}$K$-sat problem \cite{Ksat}. This
problem deals with $N$ Boolean variables represented by $\sigma_i=\pm 1$. There are $A$ clauses
with $K$ literals per clause among the Boolean variables or their negations, e.g. for $K=3$: \
$R=\{\,\sigma_i$ OR
$\sigma_j$ OR NOT $\sigma_k\,\}$. A clause $R_\mu(\{\sigma\})$ can be represented as
\begin{equation}
\xi_i^\mu=\left\{\begin{array}{rrcl}1 & \;\;{\rm if} &\;\sigma_i & \; {\rm in}\;\; R_\mu\\
-1 & \;\;{\rm if} & \;\;{\rm NOT} \,\sigma_i & \; {\rm in}\; \;R_\mu\\
0 & &\;\;{\rm else}\, .\\
\end{array}\right. 
\end{equation}
The clause $R_\mu(\{\sigma\})$ is not fulfilled if \ $\displaystyle
\sum_{i=1}^N\xi_i^\mu\sigma_i=K$, and a cost function 
\begin{equation}
H=\sum_{\mu=1}^A \delta\Big(\sum_{i=1}^N\xi_i^\mu\sigma_i-K\Big) \qquad {\rm or} \qquad
H=\sum_\mu \Big(\sum_{i=1}^N\xi_i^\mu\sigma_i-K\Big)^2
\end{equation}
may be used. Finding solutions of the $K$-sat problem, again \index{simulated annealing}simulated
annealing can be used, and finding a ground state with $H=0$  means that the corresponding choice
of the Boolean variables satisfies all clauses. The formal similarity to the Hopfield model is
obvious. For random clauses $\xi_i^\mu$ a critical
$\alpha_c(K)$ exists, such that for $\alpha=A/K<\alpha_c(K)$ solutions can typically be found by
the polynomial simulated annealing procedure.

\subsection{Minority game}

Methods, developed for disordered systems, have also been applied to various problems in economy,
especially concerning financial markets. Activities of this kind are usually subsummized under
the notion \index{econophysics}econophysics. As an example the \index{minority game}minority game
\cite{Minor} is discussed. This is a repeated game, where
$N$ agents have to decide which of two actions, e.g. buy or sell, to take. The goal is to be in
the minority. Each agent knows the outcome of the last $L$ games and can select among two
strategies, where the set of strategies is different for each agent. For each new game the agent
selects the strategy which would have been most successful over the last $L$ games. Again this
problem can be mapped onto the dynamics of an Ising spin system with quenched disorder, the
disorder being the random set of strategies for each agent. Depending on $N$ and $L$ freezing
transitions are found.\\

\section{Dynamics of the $p$-Spin Interaction Spin-Glass}\index{p-spin-glass}
\subsection{Soft Spin Models}

In the previous part of this lecture we have seen many different examples of disordered systems
exhibiting transitions to glassy dynamics and off equilibrium behaviour. Some of these systems
are based on \index{Glauber dynamics}Glauber dynamics with transition probabilities among a
discrete set of states. Glauber dynamics is very appropriate for Monte-Carlo simulations, but
difficult for analytic investigations \cite{Ho92}. In the following we therefore investigate a
model based on continuous degrees of freedom. In equilibrium the resulting equations of motion 
are identical to those derived from mode coupling theory for glasses \cite{Sch03,MCT}. The model
is, however, more general in the sense that it allows to study off equilibrium properties and
aging \cite{aging}.

Actually we investigate a whole class of models. The degrees of freedom are $N$ 
scalar real variables $\phi_i$ representing soft spins or other variables. The energy
(Hamiltonian) is
\begin{equation}\label{oygn}
H=\sum_i U(\phi_i)-\sum_i h_i\phi_i-\sfrac12 \sum_{i,j}J_{ij}\phi_i\phi_j
-\sfrac{1}{3!}\sum_{i,j,k}J_{ijk}\phi_i\phi_j\phi_k - \cdots
\end{equation}
where $U(\phi)$ is some local potential constraining the distribution of the $\phi_i$. Ising
spins for instance can be approximated by choosing $U(\phi)$ as deep double well potential
with minima at $\phi=\pm 1$, e.g. $U(\phi)=\ol{U}\,(1-\phi^2)^2$. For a \index{spherical
model}spherical model
$U(\phi)=\sfrac12\mu\phi^2$ with $\mu$ such that $\av{\phi^2}=1$.

We allow for multi spin interactions $J_{ijk}\cdots$ involving two and more spins or particles.
The interactions are Gaussian random variables with
\begin{eqnarray}\label{btes}
\ol{J_{ij}}=0 \hspace*{14.5mm}\ol{J_{ij}J_{kl}}&=&\sfrac12\{\delta_{i,k}\delta_{j,l}
+\delta_{i,l}\delta_{j,k}\}\,W^{\{2\}}_{ij}\nonumber\\
\ol{J_{ijk}}=0 \hspace*{10mm}\ol{J_{ijk}J_{lmn}}&=&\sfrac{1}{3!}
\{\delta_{i,l}\delta_{j,m}\delta_{k,n}+\cdots\}\,W^{\{3\}}_{ij}\\
\cdots\hspace*{24.8mm}\cdots &&\nonumber
\end{eqnarray}
For real spin-glasses the range of the interactions is finite, and one may choose
$W^{(2)}_{ij}=W$ if site {i} and {j} are next neighbours and $W^{(p)}_{\cdots}=0$ else
(Edwards-Anderson model \cite{EA75}). For the \index{SK-model}SK-model
(Sherrington-Kirkpatrick \cite{SK75}) on the other hand interactions with unlimited range are
used. Models with long ranged interactions (or models in infinite dimensions) have the advantage
that mean field theory yields the exact solution. For other problems like neural networks or
combinatorial optimisation problems space has no meaning and interactions among all elements are
appropriate. The strength of the interactions has to scale with the number $N$ of elements in
order to allow for a non trivial thermodynamic limit $N\to\infty$
\begin{equation}\label{htev}
W^{(p)}_{\cdots}=N^{-p/2}\,W_p.
\end{equation}
Models of this kind are often referred to as \index{mean field model}mean field models.

\subsection{Replica Theory}

Equilibrium properties can be derived from the free energy averaged over the disorder
\begin{equation}
F(T,h)=-k_B \, T \, \ol{\ln\,Z(T,h,J)}^J.
\end{equation}
The problem is to perform the average of the logarithm of the partition function instead of
averaging the partition function itself and taking the logarithm afterwards. The 
\index{replica trick}replica trick \cite{Parisi-rev,SK75} uses the identity
\begin{equation}
\ln(x)=\lim_{n\to 0}\Big(x^n-1\Big) .
\end{equation}
For integer $n$ the partition function $Z^n$ may be computed by replicating the system $n$ times
\begin{equation}
Z^n=\prod_{\alpha=1}^n\int \d\phi_1^\alpha\cdots\d\phi_N^\alpha\,
\e^{-\beta \sum_\alpha  H(\phi^\alpha)}.
\end{equation}
This quantity can now be averaged easily over the gaussian disorder. Since the disorder is the
same for each of the replica, the resulting effective Hamiltonian
\begin{equation}
-\beta{\cal H}=-\beta\sum_\alpha\ol{H(\phi_\alpha)}^J+\sfrac12\beta^2\sum_{\alpha,\beta}
\Big\{\ol{H(\phi_\alpha)H(\phi_\beta)}^J-\ol{H(\phi_\alpha)}^J\,\ol{H(\phi_\beta)}^J\Big\}
\end{equation}
couples different replica. The partition function $Z^n$ can be evaluated for mean field
models introducing the order parameters
\begin{equation}
q_{\alpha,\beta}=\ol{\av{\phi_i^\alpha,\phi_i^\beta}}^J.
\end{equation}
The problem is symmetric with respect to permutations of the replica. At high temperatures
$T>T_c$ the order parameters are symmetric with respect to permutations as well, i.e.
\begin{equation}
q_{\alpha,\beta}=q_0\delta_{\alpha,\beta}+q\,\Big(1-\delta_{\alpha,\beta}\Big). 
\end{equation}
At low temperatures $T<T_c$ two different \index{replica symmetry breaking}replica symmetry
breaking schemes have been proposed  \cite{Parisi-rev}, a one step symmetry breaking and
alternatively a hierarchical symmetry breaking in $p$ steps with $p\to\infty$, the so called full
symmetry breaking. At the end the limit $n\to 0$ has to be taken extrapolating the results
obtained originally for integer $n$ to real $n$. The results for the low temperature state with
full symmetry breaking are interpreted in terms of a state composed of distinct pure states with
ultrametric organisation with respect to their overlap. Pure states
are interpreted as regions of phase space, valleys, surrounded by barriers of infinite hight.
Within each valley the system is in equilibrium. A corresponding interpretation is obtained from
the following investigation of dynamics. For further details on the replica theory and results
see \cite{Parisi-rev,Sch03}.

\subsection{Langevin Dynamics and Path Integrals}

An investigation of dynamics yields additional information about the temporal behaviour of
correlations and especially about the critical slowing down at the 
\index{freezing transition}freezing transition.
Below the transition temperature $T<T_c$ signatures of non equilibrium are expected, in particular
\index{aging}aging.

Specifying a Hamiltonian for a classical system does not specify its dynamics uniquely.
The dynamics has, however, to be chosen such that the equilibrium state \  $\sim\e^{-\beta H}$ is
stationary. The simplest form is a Langevin equation \index{Langevin dynamics} corresponding to
the overdamped motion of particles or spins
\begin{equation}\label{nmfe}
\frac{\d\phi_i(t)}{\d t}=-\frac{\delta H(\{\phi(t)\})}{\delta \phi_i(t)}+\eta_i(t)
=F_i(\{\phi(t)\})+\eta_i(t)
\end{equation}
where the friction constant has been put to 1. The fluctuating forces, representing a heat bath, 
are assumed to be Gaussian distributed random variables with
\begin{equation}
\av{\eta_i(t)}=0 \qquad\quad 
\av{\eta_i(t)\eta_j(t')}=2\,T\,\delta_{i,j}\,\delta(t-t').
\end{equation}
In contrast to the quenched disorder of the interactions $J_{\cdots}$, the fluctuating forces
$\eta_i(t)$ depend on time. Averages over the $J_{\cdots}$ and the thermal motion, are
distinguished by different notations, $\ol{\cdots}^J$ and $\av{\cdots}$, respectively.

It is convenient, especially in view of the average over the quenched disorder, to introduce a
\index{path integral representation}path integral representation \cite{BJW76}. In the following
the derivation is outlined, dropping the site index $i$ for a moment. Let
\begin{equation}
\phi(t)=\phi(t;\{\eta\};\phi_0)
\end{equation}
be solution of \req{nmfe} with initial condition $\phi(t_0)=\phi_0$. The thermal average of a
product of $\phi$-variables at different time is then
\begin{equation}
\av{\phi(t)\phi(t')\cdots}=\!\int\!\d\phi_0P_0(\phi_0)\!\int\!\!\!\int_{t_0}\!\!{\cal D}\{\eta\}\,
\e^{{\cal W}(\{\eta\};t_0)}\phi(t;\{\eta\};\phi_0)\,\phi(t';\{\eta\};\phi_0)\cdots
\end{equation}
with
\begin{equation}
{\cal W}(\{\eta\};t_0)=-\sfrac1{4T}\int_{t_0}\d t\,\eta(t)^2.
\end{equation}
$P_0(\phi_o)$ is some distribution of initial values.

Instead of integrating over the fluctuating forces $\eta(t)$, it would be more convenient to have
a path integral over the dynamical variables $\phi(t)$. This can be done by introducing
imaginary auxiliary variables $\hat\phi(t)$, and then
\begin{equation}
\av{\phi(t)\phi(t')\cdots}=\!\int\!\d\phi_0P_0(\phi_0)\!\int\!\!\!\int_{t_0}\!\!
{\cal D}\{\hat\phi,\phi,\eta\}\,
\e^{{\cal W}(\{\hat\phi,\phi,\eta\};t_0)}\phi(t)\,\phi(t')\cdots
\end{equation}
with
\begin{equation}\label{krso}
{\cal W}(\{\hat\phi,\phi,\eta\};t_0)=-\int_{t_0}\!\d t\,
\Big\{\sfrac1{4T}\eta(t)^2+\hat\phi(t)\Big[\dot\phi(t)-F(\phi(t))-\eta(t)\Big]\Big\} .
\end{equation}
The path integral over the auxiliary imaginary variables $\hat\phi(t)$ can be viewed as a product
of $\delta$-functions insuring the fulfilment of the equation of motion, \req{nmfe}, 
at all time.

The final step is to integrate over the fluctuating forces $\eta(t)$, which
can easily be done by completing the square in \req{krso}, and now
\begin{equation}\label{ltde}
\av{\phi(t)\phi(t')\cdots}=\!\int\!\d\phi_0P_0(\phi_0)\!\int\!\!\!\int_{t_0}\!\!
{\cal D}\{\hat\phi,\phi\}\,
\e^{{\cal W}(\{\hat\phi,\phi\};t_0)}\phi(t)\,\phi(t')\cdots
\end{equation}
with
\begin{equation}\label{rski}
{\cal W}(\{\hat\phi,\phi\};t_0)=\int_{t_0}\!\d t\,
\Big\{T\hat\phi(t)^2-\hat\phi(t)\Big[\dot\phi(t)-F(\phi(t))\Big]\Big\}.
\end{equation}
The path integral extends over real functions $\phi(t)$ with initial condition
$\phi(t_0)=\phi_0$, and imaginary functions $\hat\phi(t)$ with unrestricted initial conditions.
The path integral goes over functions in the interval $t_0<t<t_1$ with $t_1$ greater than the
latest time argument in the expectation value \req{ltde}.

The auxiliary fields have a physical interpretation. Allowing for a time dependent
external field $h(t)$ in \req{oygn}, response functions can be expressed as expectation values
of the original and auxiliary fields, e.g.
\begin{equation}\label{tfoj}
\frac{\delta \av{\phi(t)}}{\delta h(t')}=\av{\phi(t)\hat\phi(t')}.
\end{equation}
The expectation value on the right hand side is of the form \req{ltde}. The auxiliary fields are
therefore denoted as response fields. 

Response functions have to be causal, i.e. \req{tfoj} has to vanish for $t'\ge t$. This is in
accordance with the Ito calculus for the stochastic equation of motion \req{nmfe}, assuming
that the action of the forces is retarded \cite{BJW76,CHS93}.

The formulation given above is completely general in the sense that it is not restricted to
equilibrium or small deviations from equilibrium. In general correlation- and response-functions
are not related. In equilibrium, however, they have to obey
\index{fluctuation-dissipation-theorem}fluctuation-dissipation-theorems (FDT)
\begin{equation}\label{gepi}
\beta\,\frac{\d}{\d t'}\,q(t,t')=r(t,t')
\end{equation}
where correlation- and response-function are defined as
\begin{equation}
q(t,t')=\av{\phi(t)\phi(t')} \qquad\quad r(t,t')=\av{\phi(t)\hat\phi(t')}.
\end{equation}

A sketch of the proof is as follows: The first step is to show that for equilibrium initial
condition
\begin{equation}\label{nges}
P_0(\phi_0)=Z^{-1}\e^{-\beta H(\phi_0)}
\end{equation} 
the initial time is arbitrary as long as it is earlier
than the earliest time in expectation values of the type \req{ltde} or \req{tfoj}. Assume that
the initial distribution at some time $t_0'<t_0$ is given by \req{nges}. For $t'_0<t<t_0$ one
replaces $\hat\phi(t)\to \hat\phi(t)+\beta\dot\phi(t)$ in the path integral \req{ltde} and in the
action
\req{rski}. This results in
\begin{eqnarray}\label{hrao}
{\cal W}(\{\hat\phi,\phi\};t_0)-{\cal W}(\{\hat\phi,\phi\};t'_0)
&=&\int_{t'_0}^{t_0}\!\d t\,
\Big\{T\hat\phi(t)^2+\hat\phi(t)\Big[\dot\phi(t)+F(\phi(t))\Big]\Big\}\quad\nonumber\\
&-&\beta H(\phi(t_0))+\beta H(\phi(t'_0)).
\end{eqnarray}
The last two terms in \req{hrao} originate from integrating a contribution \linebreak
$\beta\dot\phi(t)F(\phi(t))=-\beta \d H(\phi(t))/\d t$ which appears if the above substitution is
performed. Evaluating the path integral over $\phi(t)$ or actually $\dot\phi(t)$ for $t'_0<t<t_0$
the only contributions come from $\hat\phi(t)=0$. The last two terms in \req{hrao} replace the
initial equilibrium condition 
\req{nges} at $t'_0$ by the corresponding one at $t_0$. If the system is ergodic, the limit
$t'_0\to-\infty$ may be taken and, as a consequence, the initial condition becomes irrelevant.

Consider now a small change of the external field $\delta h(t)=\delta h_0$ for $t<t_0$ and $\delta
h(t)=0$ for $t>t_0$. Assuming equilibrium 
\begin{equation}\label{frsa}
\frac{\delta\av{\phi(t)}}{\delta h_0}=\int_{-\infty}^{t_0}\!\!\d t' \,r(t,t')=
\beta q(t,t_0).
\end{equation}
For the   first expression the initial time $t'_0\to-\infty$ is used. The second is
obtained by differentiating the equilibrium initial condition at $t_0$ with respect to $h$. The
FDT is obtained by differentiation both expressions of \req{frsa} with respect to $t_0$.

Deriving this result no explicit time dependence of the Hamiltonian is allowed and the system
has to be ergodic.

\subsection{Average over Quenched Disorder}\index{quenched disorder}

As pointed out already in the context of the replica theory, the calculation simplifies
considerably for mean field models with long ranged interactions, specified in \req{htev}. In the
following we investigate models of this kind \cite{SZ82,Ho84,Kir87,CHS93}.

Provided the initial condition does not depend on the
disorder, the average over the interactions $J^{(p)}_{\cdots}$ can easily be performed
using the general relation 
\begin{equation}
\ol{\e^{ax}}=\e^{a\ol{x}+\frac12\ol{(x-\ol{x})^2}}
\end{equation}
valid for Gaussian distributions.
This yields, for the Hamiltonian \req{oygn} and the disorder specified
in \req{btes} and \req{htev}, the effective action
\begin{eqnarray}
\ol{{\cal W}(\{\hat\phi,\phi\})}^J&=&\int\!\!\d t\sum_i\Big\{T\hat\phi_i(t)^2
-\hat\phi_i(t)\Big[\dot\phi_i(t)-h-U'\big(\phi_i(t)\big)\Big]\Big\}\nonumber\\
&+&\int\!\!\d
t\!\!\int^t\!\!\d t'\sum_i\Big\{\hat\phi_i(t)
W'\Big(\sfrac1N\sum_j\phi_j(t)\phi_j(t')\Big)\hat\phi_i(t')\\
&&\hspace*{1.5mm}+\hat\phi_i(t)\sfrac1N\sum_j\phi_j(t)
W''\Big(\sfrac1N\sum_k\phi_k(t)\phi_k(t')\Big)\hat\phi_j(t')\phi_i(t')\Big\}\quad\nonumber
\end{eqnarray}
with
\begin{equation}\label{jrso}
W(x)=\sum_p\sfrac1{p!}\,W^{(p)}\,x^p.
\end{equation}
Correlation- and response- functions calculated with this action are now disorder averaged
quantities
\begin{equation}
q(t,t')=\sfrac1N\sum_i\ol{\av{\phi_i(t)\phi_i(t')}}^J 
\qquad\quad r(t,t')=\sfrac1N\sum_i\ol{\av{\phi_i(t)\hat\phi_i(t')}}^J.
\end{equation}

\subsection{Dynamic Mean Field Theory for Spherical Models}
\index{dynamic mean field theory}

Computing local correlation- and response-function, i.e. averages involving $\hat\phi_i$ and
$\phi_i$ at a single site $i$ only, a saddle point evaluation of the corresponding path integral
is possible. The local time dependent functions are obtained from an effective single site action
\begin{eqnarray}
\ol{{\cal W}_{\rm eff}(\{\hat\phi_i,\phi_i\})}^J&=&\int\!\!\d t\Big\{T\hat\phi_i(t)^2
-\hat\phi_i(t)\Big[\dot\phi_i(t)-h-U'\big(\phi_i(t)\big)\Big]\Big\}\nonumber\\
&+&\int\!\!\d
t\!\!\int^t\!\!\d t'\Big\{\hat\phi_i(t)\,
W'\Big(q(t,t')\Big)\,\hat\phi_i(t')\\
&&\hspace*{3.5mm}+\hat\phi_i(t)\,r(t,t')
W''\Big(q(t,t')\Big)\,\phi_i(t')\Big\}\quad\nonumber
\end{eqnarray}
which has to be determined  selfconsistently. This is exact for mean field models in the
thermodynamic limit $N\to\infty$. 

Compared to conventional mean field theory, e.g. for magnets, the present theory is much richer
because the order parameters are the correlation- and response-functions and therefore functions
of two time arguments.

For spherical models with $U(x)=\sfrac12 \mu x^2$ the effective action is quadratic, which
simplifies the calculation further. In particular it allows to write down the resulting dynamical
mean field equations in the following closed form \cite{Kir87,CHS93}
\begin{eqnarray}\label{hfre}
\Big(\frac{\d}{\d t}+\mu(t)\Big)\,q(t,t')&=&h\,m(t')+\int_{t'}^t\d s\,K(t,s)q(s,t')
\nonumber\\
&+&\int_{t_0}^{t'}\d s \,\Big\{M(t,s)r(t',s)+K(t,s)q(t',s)\Big\}\\
\Big(\frac{\d}{\d t}+\mu(t)\Big)\,r(t,t')&=&\int_{t'}^t\d s\,K(t,s)r(s,t')\\
\Big(\frac{\d}{\d t}+\mu(t)\Big)\;\,m(t)\;\,&=& h+\int_{t_0}^t\d s\,K(t,s)\,m(s)
\end{eqnarray}
with
\begin{eqnarray}\label{bgsx} 
K(t,t')&=&W''\big(q(t,t')\big)\,r(t,t')\nonumber\\
M(t,t')&=&W'\,\big(q(t,t')\big).
\end{eqnarray}
The spherical constraint $q(t,t)=1$ yields
\begin{equation}\label{nfeu}
\mu(t)=h m(t)+T+\int_{t_0}^t\d s\,\Big\{K(t,s)q(t,s)+M(t,s)r(t,s)\Big\}.
\end{equation}

The above dynamical mean field equations are a set of coupled non linear
integro-differential-equations for the correlation- and response-functions. Initial conditions
are $q(t,t)=1$ and $r(t,t)=1$. It should be stressed that the above equations do not require
equilibrium. They are therefore suited to deal with off equilibrium properties and aging. In
general the correlation- and response-functions depend on both time arguments $t$ and $t'$ and
not only on the difference $t-t'$.

\subsection{Equilibrium Dynamics in the Ergodic Phase}\index{ergodic phase}

Above the freezing temperature $T_c$, in equilibrium, the correlation- and res\-ponse-functions
depend on the difference
$t-t'$ only and they obey
\index{fluctuation-dissipation-theorem}fluctuation-dissipation-theorems, \req{gepi}. The above
equations of motion simplify considerably. The \index{memory terms}memory terms obey
fluctuation-dissipation-theorems as well
\begin{equation}
K(t)=-\beta \, \dot M(t)
\end{equation}\\[-8mm]
and \req{hfre} reads
\begin{equation}\label{gtem}
\Big(\frac{\d}{\d
t}+\ol{\mu}\Big)q(t)=h\,\ol{m}+\beta\Big\{M(t)-M(\infty)q(\infty)\Big\}
-\beta\!\!\int_0^t\!\!\d s\,
\dot M(t-s)q(s)\quad
\end{equation}
with $M(t)=W'\big(q(t)\big)$. 
Magnetisation $\ol{m}=m(t\!\to\!\infty)$ and $\ol{\mu}=\mu(t\!\to\!\infty)$ are given
by
\begin{equation}\label{hrpo}
\ol{\mu}\,\ol{m}=h+\beta\,\big\{W'(1)-W'(\ol{q})\big\}\,\ol{m}
\end{equation}
with $\ol{q}=q(t\!\to\!\infty)$, and
\begin{equation}\label{bnek}
\ol{\mu}=h\ol{m}+T+\beta\,\big\{W'(1)-\ol{q}\,W'(\ol{q})\big\}.
\end{equation}

\begin{figure}[ht]
\centering
\includegraphics[width=8cm]{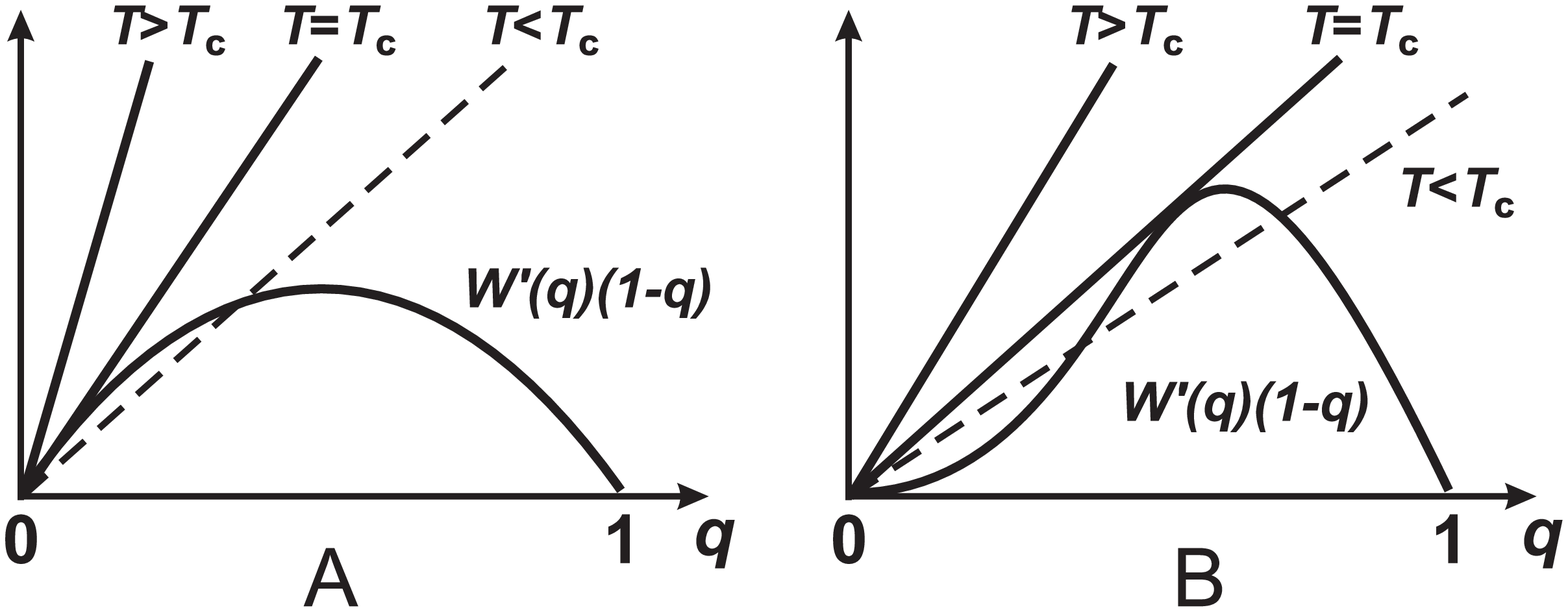} 
\caption{Stability criterium \req{nrsy}  for $h=0$. \ (A):  $p=2$ \ (B):  $p>2$. The left hand
side is plotted for various temperatures. For $T<T_c$ the \index{stability criterium}stability
criterium is violated.}
\label{Stability} 
\end{figure}

For $h=0$ one has \ $\ol{m}=\ol{q}=0$ and \ $\ol{\mu}=T+\beta\,W'(1)$. The following discussion
will be restricted to $h=0$. The general case with $h\ne0$ is discussed in \cite{CHS93} and
results will be shown later.\\
For general $t$, \req{gtem} can be solved numerically. Since we are dealing here with a purely
dissipative system, \ $\dot q(t)\le 0$ is required for all $t$. Replacing $q(s)$ by $q(t)$ in the
integrand of \req{gtem} the following inequality is obtained
\begin{equation}
0\le -\dot q(t)\le q(t)
-\beta\,W'\big(q(t)\big)\,\big(1-q(t)\big)
\end{equation}
or
\begin{equation}\label{nrsy}
T\,q(t)\ge W'\big(q(t)\big)\,\big(1-q(t)\big).
\end{equation}
Left and right hand side of \req{nrsy} are shown in Fig.\ref{Stability}. 
\\
\newpage
\noindent Assuming $W(q)\sim q^p$ for $q\to0$ two cases have to be distinguished:\\[-6mm]
\begin{itemize}
\item[A)]  $p=2$: For $h=0$ a second solution $q_c$ branches off for $T<T_c$. The stability
criterium is, however, violated for $q(t)<q_c$. This means that a phase transition into an off
equilibrium phase has to take place at $T=T_c$. Since this solution branches off continuously
from $\ol{q}=0$ at $T_c$ this transition is referred to as 
\index{continuous transition}continuous transition. For
$h\ne0$ the straight line representing the left hand side cuts the $q\!=\!0-$axis at some
negative value and there is no transition in finite field \cite{Zip00}.
\item[B)]  $p>2$: \ For $T\!>\!T_c$  again a single solution $\ol{q}=0$ exists. At $T_c(h)$ a new
solution $q_c$ fulfilling
\begin{equation}\label{hrsa}
W'\big(q_c\big)\,\big(1-q_c\big)/q_c=T_c
\end{equation}
shows up. This means that the resulting phase transition is discontinuous.
\index{discontinuous transition}
For $T=T_c$ and $q=q_c$ the slope of the left and right hand side of \req{nrsy} has to
be the same resulting in
\begin{equation}\label{nrop}
T_c=W''(q_c)\,(1-q_c)-W'(q_c).
\end{equation}
The two Eqs.(\ref{hrsa}) and (\ref{nrop}) determine $T_c$ and $q_c$.
\end{itemize}
\begin{figure}[ht]
\centering
\includegraphics[width=6cm]{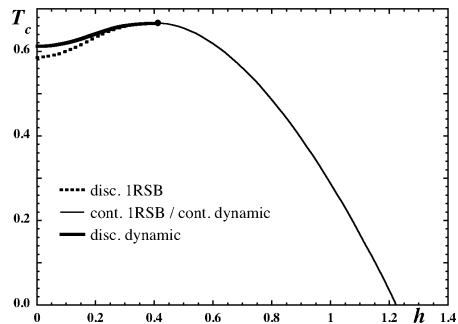} 
\caption{Phase diagram of the spherical $p$-spin interaction spin-glass \cite{CHS93}.}
\label{PhaseD} 
\end{figure}
This discussion can be extended to $h\ne0$ \cite{CHS93}.
The resulting \index{phase diagram}phase diagram is shown in Fig.\ref{PhaseD} for the
$p=3$-spin-glass with
$W(q)=\sfrac{1}{3!}q^3$.

For $h<0.41$ the transition is discontinuous. The transition temperatures $T_c(h)$ 
obtained from dynamics is higher than the one found in replica theory. This surprising result
might be explained by the proposal that the dominant states counted for in replica theory are not
reached by the dynamics investigated \cite{Bar96}.

For $h>0.41$ the transition is continuous and both theories give the same transition temperature.

\begin{figure}[ht]
\centering
\includegraphics[width=6cm]{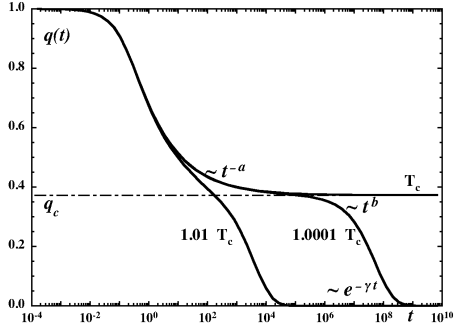} 
\caption{Correlation function $q(t)$ for $T\ge T_c$ \cite{CHS93}.}
\label{q(t)>Tc} 
\end{figure}

The results of a numerical integration of \req{gtem} for $h=0$ is shown in Fig.\ref{q(t)>Tc} for
temperatures slightly above and at the critical temperature $T_c$. 
The asymptotic behaviour
$q(t)$ close to $T_c$ can be analysed by identifying appropriate scaling functions for different
regions in time and by matching \cite{MCT,Kir87,CHS93}. This resembles very much the
\index{crossover scaling}crossover scaling analysis near multi critical points \cite{MultCrit}.
From Fig.\ref{q(t)>Tc} one can read off two characteristic temperature dependent time scales: A
\index{plateau}plateau time $\tau_p(T)$ can be defined as $q(\tau_p)=q_c$ and a second time scale
$\tau_a(T)$ characterising the final decay. It may be defined as $q(\tau_a)=\sfrac12 q_c$.\\
\\
For $t\ll \tau_p(T)$ the correlation-function approaches some universal function 
$q(t)\to\hat q_o(t)$ as $T\to T_c$. For $1\ll t\ll \tau_p(T)$ \req{gtem} is solved by
\begin{equation}\label{bgeu}
\hat q_o(t)\to q_c+c_o\,t^{-a}.
\end{equation}
The dynamical critical exponent $a$ is solution of
\begin{equation}\label{ouyd}
\frac{\Gamma(1-a)^2}{\Gamma(1-2a)}=\frac{(1-q_c)W'''(q_c)}{2\,W''(q_c)}.
\end{equation}
This result is obtained by inserting \req{bgeu} into \req{gtem} with $T=T_c$. 
For $t\!\to\!\infty$ the leading contribution is ${\cal O}(1)$. The next to leading contributions
${\cal O}(t^{-a})$ yield \req{hrsa} and \req{nrop} respectively. Collecting terms ${\cal
O}(t^{-2a})$ \req{ouyd} is obtained.

The scaling function ansatz for $t\sim \tau_p$ is
\begin{equation}\label{hfsw}
q(t)\approx q_c+\tau_p^{-a}\hat q_p(t/\tau_p).
\end{equation}
The scale factor $\tau_p^{-a}$ follows from matching \req{bgeu} and \req{hfsw} at some 
$1\ll t \ll \tau_p$ and $\hat q_p(\tau)\to c_o\tau^{-a}$ for $\tau\to0$. 

\newpage

The contributions ${\cal O}(T-T_c)$ to \req{bgeu} for $t\sim \tau_p$ give
\begin{equation}\label{potd}
\tau_p\sim(T-T_c)^{-1/2a}.
\end{equation}
For $\tau_p\ll t\ll \tau_a$ 
\req{gtem} is solved by
\begin{equation}\label{hjte}
\hat q_p(\tau)\to -c_p\,\tau^b
\end{equation}
with $b$ obeying
\begin{equation}\label{tydh}
\frac{\Gamma(1+b)^2}{\Gamma(1+2b)}=\frac{(1-q_c)W'''(q_c)}{2\,W''(q_c)}=
\frac{\Gamma(1-a)^2}{\Gamma(1-2a)}.
\end{equation}
This means that the two dynamical critical exponents $a$ and $b$ are not independent 
\cite{MCT,CHS93}.

The last scaling regime applies for $t\sim \tau_a$. The scaling ansatz is
\begin{equation}\label{kjts}
q(t)=\hat q_a(t/\tau_a).
\end{equation}
Matching to \req{hfsw} yields
\begin{equation}
\hat q_a(\tau)\to q_c-c_p\,\tau^b \quad {\rm for} \quad \tau\to0
\end{equation}
and
\begin{equation}\label{mnyx}
\tau_p=\tau_a^{b/(a+b)}
\end{equation}
and with \req{potd}
\begin{equation}\label{gjyo}
\tau_a\sim(T-T_c)^{-(a+b)/2ab}.
\end{equation}
This means that only a single independent \index{critical exponent}critical exponent exists. The
scaling functions 
$\hat q_o(t)$, $\hat q_p(\tau)$ and $\hat q_a(\tau)$ have to be determined numerically. 

The imaginary part of the frequency dependent \index{susceptibility}susceptibility is
\begin{equation}
\chi''(\omega)=\int_0^{\infty}\!\d t\,r(t)\,\sin(\omega t)=
\beta\,\omega\int_0^{\infty}\!\d t\,q(t)\,\cos(\omega t)
\end{equation}
where the second expression is derived from the FDT, \req{gepi}. Near $T_c$ the main
contributions are due to $t\sim 1$ and $t\sim \tau_a$, respectively. Inserting the scaling
discussed above
\begin{eqnarray}
\chi''(\omega)&=&\beta\omega\int_0^{\infty}\!\d t\,q_0(t)\,\cos(\omega
t)+\beta\sfrac{\omega}{\omega_a}\int_0^\infty\!\d\tau\,\hat
q_a(\tau)\,\cos(\sfrac{\omega}{\omega_a}\tau)\nonumber\\
&=&\chi''_o(\omega)+\chi_a''(\sfrac{\omega}{\omega_a})
\end{eqnarray}
with
\begin{equation}
\omega_a=1/\tau_a\sim (T-T_c)^{(a+b)/2ab}.
\end{equation} 
With \req{bgeu} for $\omega\to0$
\begin{equation}
\chi''_o(\omega)\to \beta c_o\Gamma(1-a)\cos(\sfrac12\pi(1-a))\,\omega^a
\end{equation}
is found. For $\omega\ll\omega_a$ the second contribution with \req{hjte} gives
\begin{equation}
\chi''_a(\omega)\to \beta c_a\Gamma(1+b)\cos(\sfrac12\pi(1+b))\,\omega^{-b}.
\end{equation}
Combining both expressions $\chi''(\omega)$ shows a minimum at a frequency
\begin{equation}
\omega_p\sim \tau_a^{b/(a+b)}\sim \tau_p^{-1}\sim (T-T_c)^{1/2a}.
\end{equation}\\[-11mm]
\begin{figure}[ht]
\centering
\includegraphics[width=5.5cm]{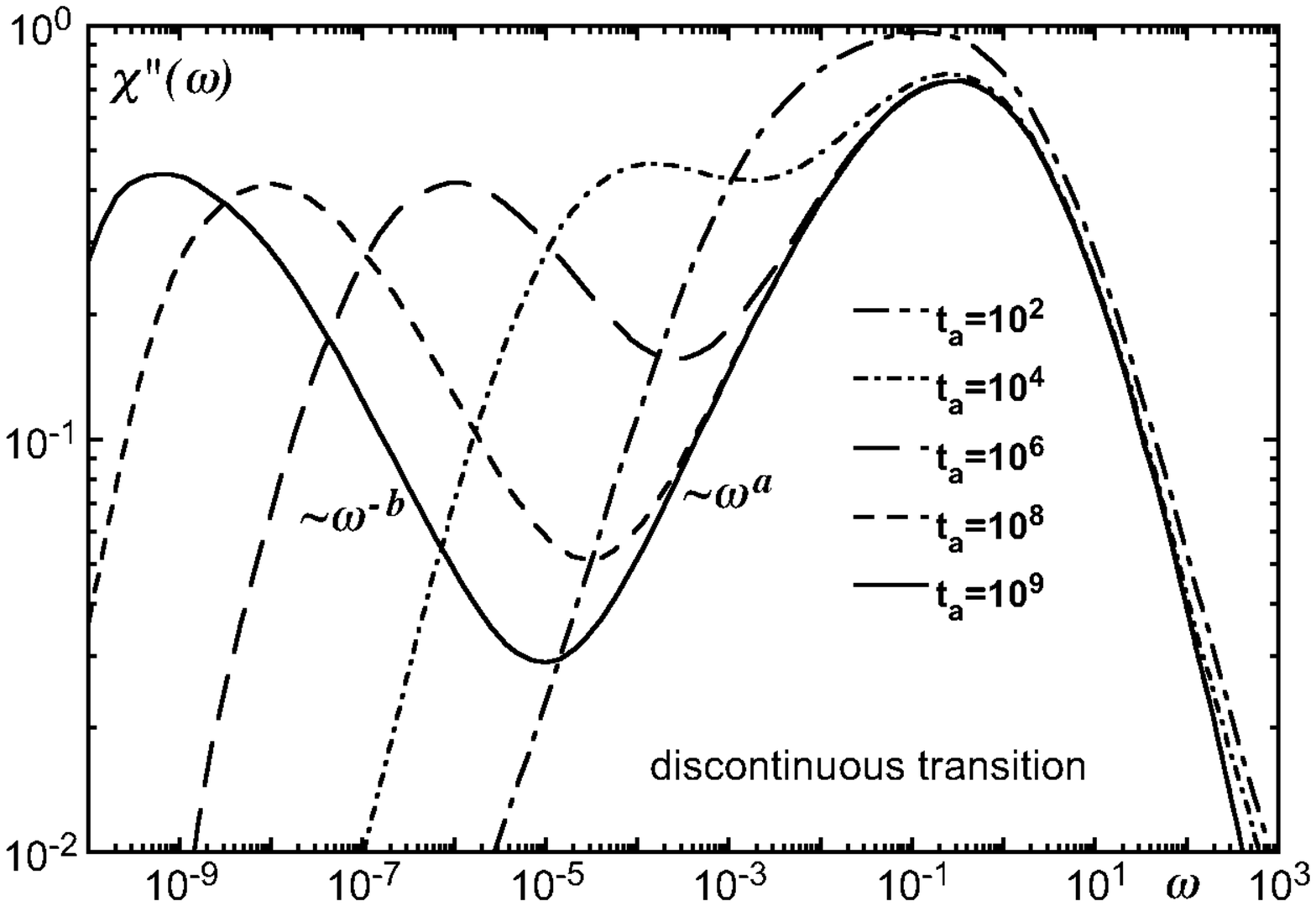} 
\includegraphics[width=5.5cm]{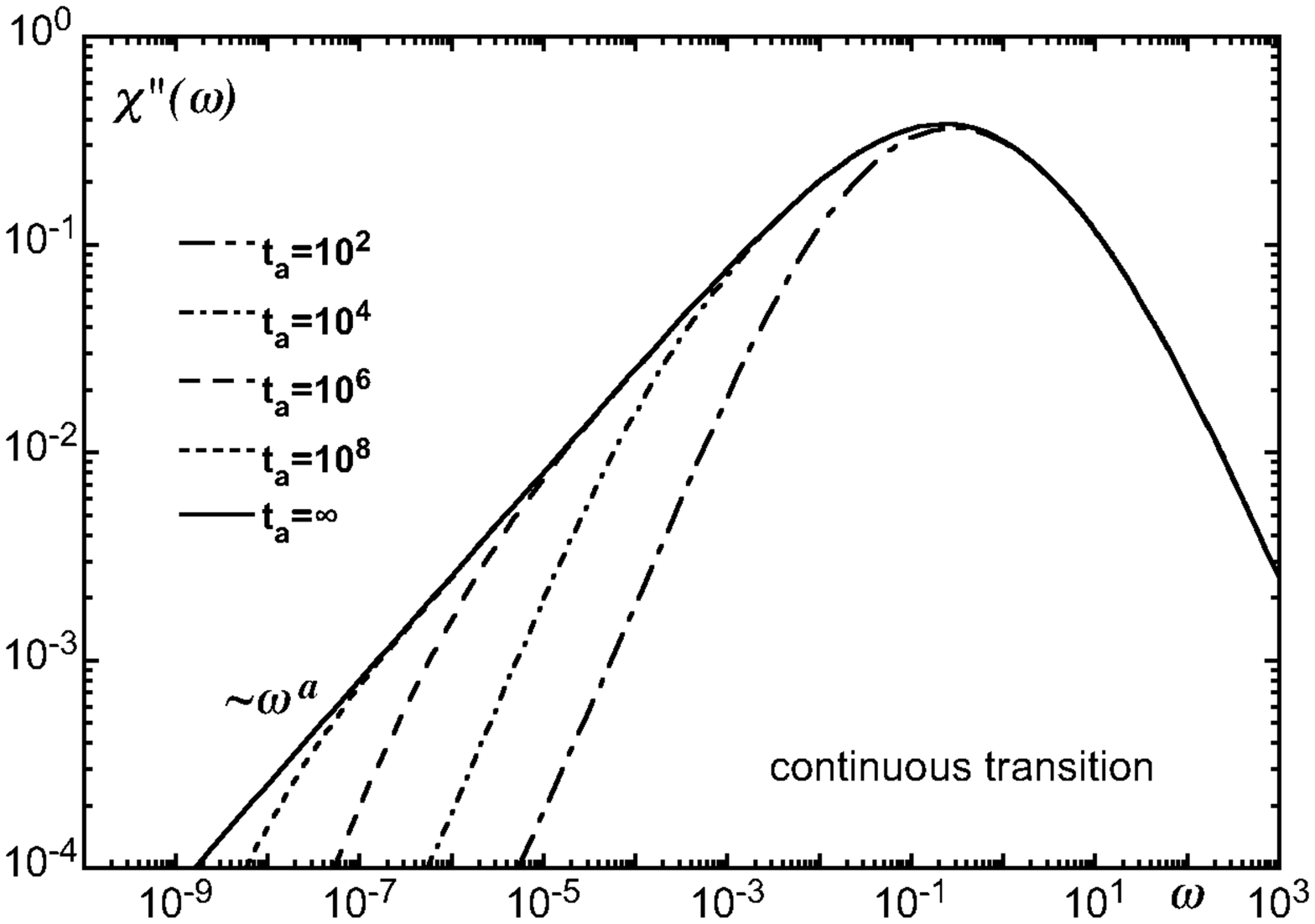} 
\caption{Susceptibility $\chi''(\omega)$ computed from the numerical solution of 
\req{gtem} for $h=0$ and various temperatures characterised by $\tau_a(T)$, \req{gjyo}. For
comparison the susceptibility for finite $h$ in the region of continuous transitions is also
shown.}
\label{X(omega)} 
\end{figure}\\
The complete \index{susceptibility}susceptibility is plotted in Fig.\ref{X(omega)}. The
similarity with the experiments on glasses, shown in Fig.\ref{CKN}a, is obvious, and indeed the
data on
$CKN$ and other glasses have been fitted to the equivalent results of \index{mode coupling
theory}mode coupling theory
\cite{MCT,CKN92,Goe99} reproducing the data quite well.

\subsection{Off Equilibrium Dynamics and Aging in the QFDT-Phase}\label{ojrb}\index{QFDT-phase}

The investigation of the dynamics for $T<T_c$ requires to specify some long time scale
$\tau_\infty$, and to consider the limit $\tau_\infty\to\infty$ eventually. There have been
several  proposals for such a time scale:\\[-5mm]
\begin{itemize}
\item[$\bullet$] For finite systems with $N$ elements the equilibration time $t_\infty$ is
finite \cite{SZ82}. For large $N$, \, $\tau_\infty \sim \e^{N^{1/3}}$ \cite{KiHo91}.\\[-3mm]
\item[$\bullet$] Relaxing interactions 
$\ol{J(t)\,J(t')}\sim \e^{(t-t')/\tau_\infty}$ \cite{Ho84}.\\[-3mm]
\item[$\bullet$] Slow cooling with $T(t)=(1-t/\tau_\infty)\,T_c$ \cite{CHS93,Ho86}.\\[-3mm]
\item[$\bullet$] Fast cooling, aging, with $\tau_\infty=t_w$ \cite{CuK93,Ho03}.
\end{itemize}
In the following we concentrate on \index{aging}aging. The system is rapidly cooled from
high temperatures to some $T<T_c$  at $t=0$. Alternatively a strong magnetic field could be
switched off at $t=0$. The requirement is, that the initial state is not correlated with the
disorder.

\newpage

\begin{figure}[ht]
\centering
\includegraphics[width=8cm]{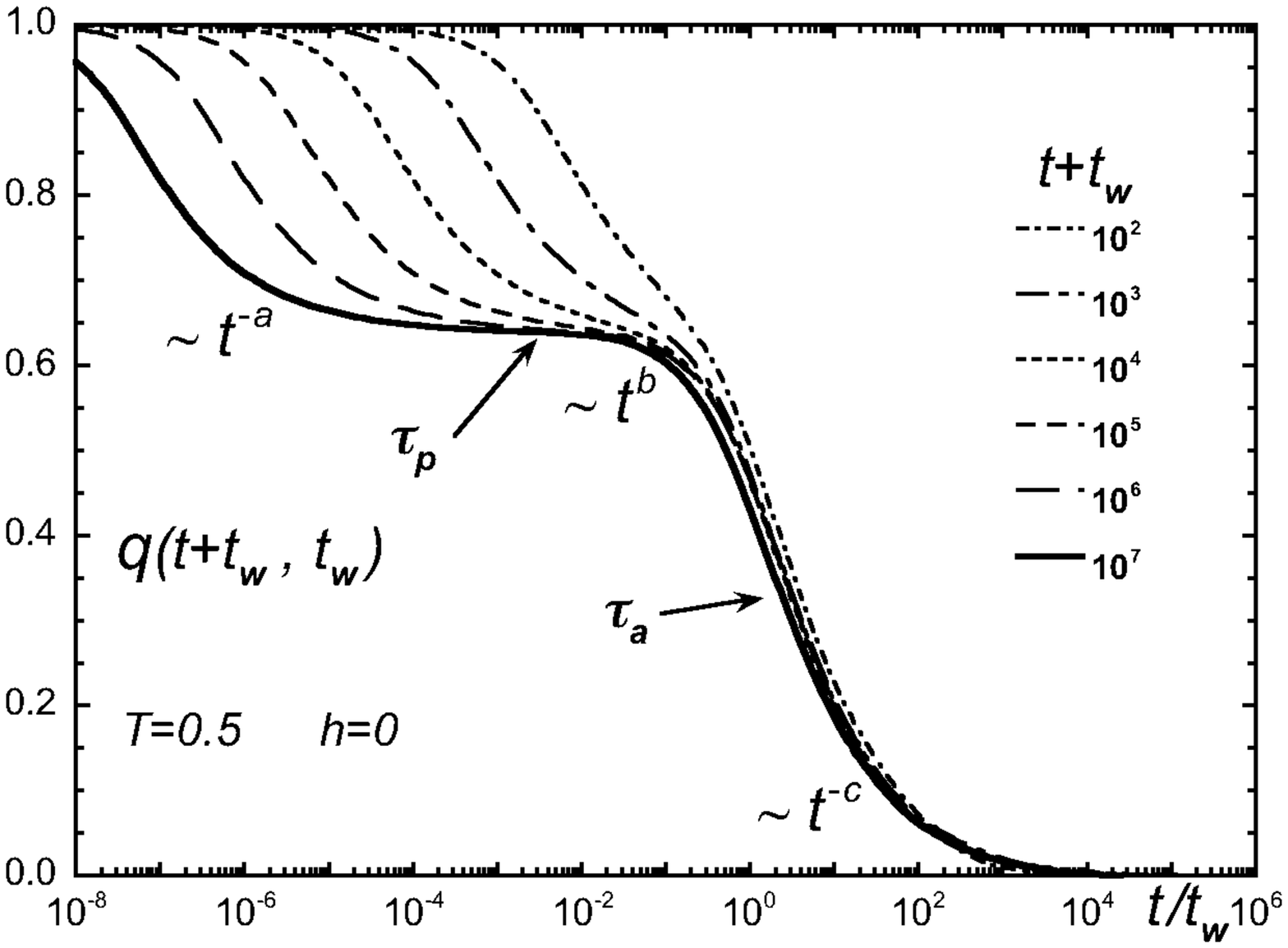} 
\includegraphics[width=8cm]{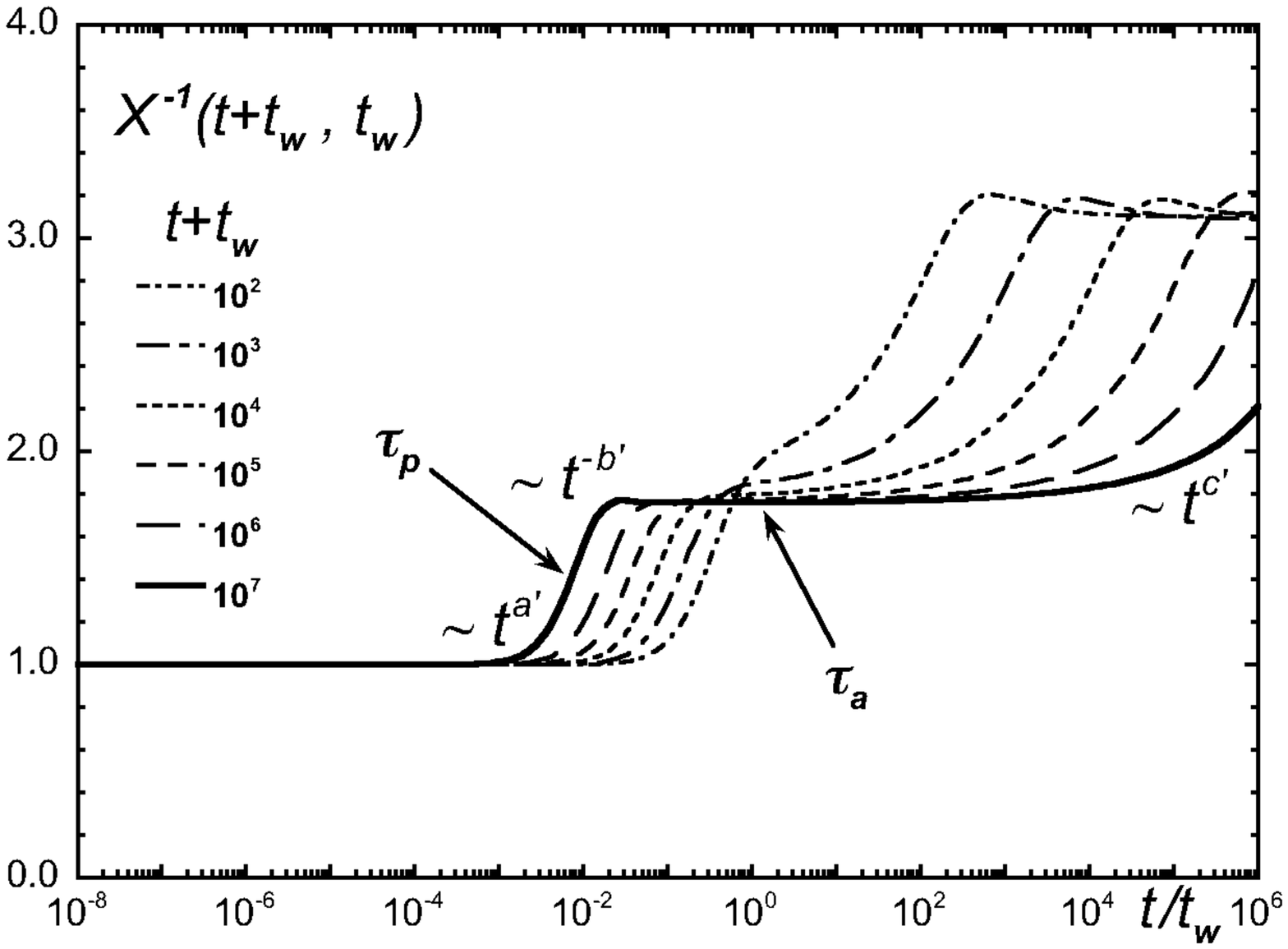} 
\caption{Correlation function $q(t+t_w,t_w)$ and FDT-violation $X(t+t_w,t_w)$ for $T<T_c$ and
$h=0$. The curves are plotted for fixed $t+t_w$ as functions of $t/t_w$. The range $t/t_w\sim
t+t_w$ corresponds to
$t_w\sim 1$.}
\label{qX(t,tw)} 
\end{figure}

Results from a numerical integration \cite{Ho86,Ho03} of the dynamical mean field equations
(\ref{hfre}-\ref{nfeu}) are shown in Fig.\ref{qX(t,tw)}.

Because the system is not in equilibrium at $T<T_c\,$, correlation- and response-functions depend
on both time variables $t$ and $t'$, and not on $t-t'$ only as in equilibrium. Furthermore
fluctuation dissipation theorems, \req{gepi}, are not fulfilled. A measure for
\index{FDT-violation}FDT-violation is 
\begin{equation}
X(t,t')=T\frac{r(t,t')}{\sfrac{{\rm d}}{{\rm d} t'}q(t,t')}
\end{equation}
which is shown in the lower part of Fig.\ref{qX(t,tw)}. This quantity can be relates to an
\index{effective temperature}effective temperature \cite{CKP97}
\begin{equation}
T_{\rm eff}(t,t')=X^{-1}(t,t')\,T.
\end{equation} 

For short time $X(t+t_w,t_w)\approx1$ is found. This means that the FDT is fulfilled, while
$q(t+t_w,t_w)$ decays from 1 to $q_c$. A distance in phase space can be defined as
\begin{equation}
D(t,t_w)=\sqrt{\sfrac12\av{\big(\phi_i(t+t_w)-\phi_i(t_w)\big)^2}}
=\sqrt{\,1-q(t+t_w,t_w)\Big.\,}.
\end{equation}
This distance grows until $q(t+t_w,t_w)\approx q_c$ is reached, and the resulting
$D_c=\sqrt{1-q_c}$ can be interpreted as the size of a typical valley in phase space. The
system apparently equilibrates within such a valley, and escapes only after some characteristic
time
$\tau_p(t_w)$. This escape is irreversible, which is indicated by the deviation of $X(t+t_w,t_w)$
from 1 starting at $t\sim \tau_p(t_w)$. 

For $t>\tau_p(t_w)$ the parameter $X(t+t_w,t_w)$ reaches a \index{plateau}plateau at some $X_c$,
which extends up to $t\gg t_w$. In this regime a modified fluctuation dissipation theorem (QFDT)
\index{QFDT} holds.
A solution of this kind was first observed in the context of learning in 
\index{neural network}neural networks exhibiting a \index{discontinuous transition}discontinuous
transition as well \cite{Ho92}. It appears to be a common feature of discontinuous freezing
transitions \cite{Ho96,CHS93}.

Investigating the asymptotic behaviour of the solutions of the dynamical mean field equations, 
Eqs.(\ref{hfre}-\ref{nfeu}), the parameters $q_c$, $\ol{q}$, $\ol{\mu}$, $\ol{m}$ and $X_c$ can
be evaluated analytically \cite{CHS93}. Especially $q_c$ is given by
$q_c=\beta\,W'(q_c)\,(1-q_c)$, which is of the form of \req{hrsa}, indicating that the solution is
marginally stable \index{marginal stability} at all temperatures $T<T_c$. 

The \index{crossover scaling}crossover scaling analysis is similar to the one discussed for
ergodic dynamics in the previous section, but the time scales $\tau_p(t_w)$ and $\tau_a(t_w)$
are now determined by the waiting time$t_w$. 

Again dynamical \index{critical exponent}critical exponents are introduced. They obey
\begin{equation}\label{btsp}
\frac{\Gamma(1-a)^2}{\Gamma(1-2a)}=X_c\,\frac{\Gamma(1+b)^2}{\Gamma(1+2b)}=
\frac{(1-q_c)W'''(q_c)}{2\,W''(q_c)}
\end{equation}
which is identical to the result \req{tydh} for $T>T_c\,$, except for the factor $X_c$. The
behaviour of $X(t+t_w,t_w)$ for $1\ll t\ll \tau_p$ and $\tau_p\ll t \ll \tau_a$, respectively, is
also ruled by power laws, as indicated in the lower part of Fig.\ref{qX(t,tw)}. The resulting
exponents fulfil \cite{Ho86,Ho03}
\begin{equation}
a'=3a+1 \qquad\quad b'=3b-1.
\end{equation}

Obviously for the existence of the \index{QFDT-phase}QFDT-phase $b'>0$ has to be fulfilled. $b'=0$
therefore marks a phase transition from the QFDT-phase ($\cal A$-phase) \index{A-phase} to a phase
with a hierarchy of long time scales ($\cal B$-phase)\index{B-phase}. Another criterium to be
fulfilled is
$X(t,t')\ge1$. This gives rise to yet another phase ($\cal C$-phase) which covers only a tiny
portion of the phase diagram.

\begin{figure}[ht]
\centering
\includegraphics[width=8cm]{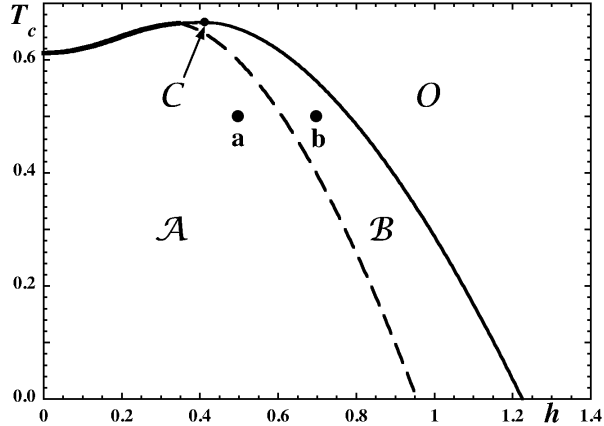} 
\caption{Phase diagram of the spherical $p\!\!=\!\! 3-$spin interaction
spin-glass \cite{Ho03}. \ \
${\cal O}$: Ergodic phase. ${\cal A}$: QFDT-phase. ${\cal B}$ and ${\cal C}$: Phases with a
hierarchy of long time scales. The points {\bf a} and {\bf b} are discussed later.}
\label{PhaseDiax}
\end{figure}
The complete \index{phase diagram}phase diagram is shown in Fig.\ref{PhaseDiax}. Qualitatively
the same phase diagram is found for the mean field model of a particle moving \index{drift of a
particle} in a correlated random potential \cite{Ho96}. 

In phase ${\cal A}$ the plateau-time $\tau_p$ and the characteristic time scale for aging $\tau_a$
are again related by \req{mnyx} with the modified relationship, \req{btsp}, among the exponents. 

For $t\sim t_w$ it has been argued  \cite{CuK93,Cug03} that
\begin{equation}
q(t+t_w,t_w)=Q\big(h(t+t_w)/h(t_w)\big)
\end{equation}
as long as $X(t+t_w,t_w)\approx X_c$. The function $h(t)$ has to be monotonic, but is otherwise
not determined. The arbitrariness in $h(t)$ resembles the invariance \index{reparametisation
invariance} postulated for the long time behaviour of the SK-Model \cite{SZ82,Ho84}.  

$Q(x)$ can be computed numerically and $Q(x)-q_c\sim (x-1)^b$ for $x\to1$.
Especially for zero field the exponent $b=1$ and
\begin{equation}
Q(x)=\frac{q_c}{x}
\end{equation}
is found. The postulate of scaling with $\tau_a(t_w)=t_w$ yields $h(t)=t^c$. The numerical
results \cite{Ho03} shown in Fig.\ref{qX(t,tw)} agree reasonably well with this form of scaling
and in particular with $c\approx 1/2$.

For the analogous case of the drift of a particle in a random potential \index{drift of a
particle} \cite{Ho96}, \ $\tau_a\sim \tau_\infty^\mu\,$, with
$\mu=1-a(3b-1)/\big(2(a+b)+a(3b-1)\big)$, has been found in the ${\cal A}$-phase. In this problem
the time scale $\tau_\infty=1/v_{\rm drift}$ plays the same role as $t_w$ for aging. Typical
values  are $\mu\sim 0.9\cdots0.95$. Scaling of the form given in \req{kjts} with $\tau_a(t_w)\sim
t_w^{\mu}$ is obtained from  $h(t)=\e^{(t^\eta-1)\,c/\eta}$ with $\eta=1-\mu$.
The numerical solutions \cite{Ho03} indicate $\eta>0$, but the longest
time investigated is not long enough, to decide on the actual value of $\eta$.

\subsection{Off Equilibrium Dynamics and Aging in the ${\cal B}$-Phase}
\label{vbrt}\index{B-phase}

We now turn to the ${\cal B}$-phase. Fig.\ref{cX(t,tw)-AB} shows
correlation-function $c(t+t_w,t_w)=q(t+t_w,t_w)-m(t+t_w)\,m(t_w)$ and
\index{FDT-violation}FDT-violation parameter $X(t+t_w,t_w)$ for field and temperature values
indicated as points {\bf a} and {\bf b} in Fig.\ref{PhaseDiax}. The qualitative
difference between ${\cal A}$- and ${\cal B}$-phase is clearly visible.
\begin{figure}[ht]
\centering
\includegraphics[width=10cm]{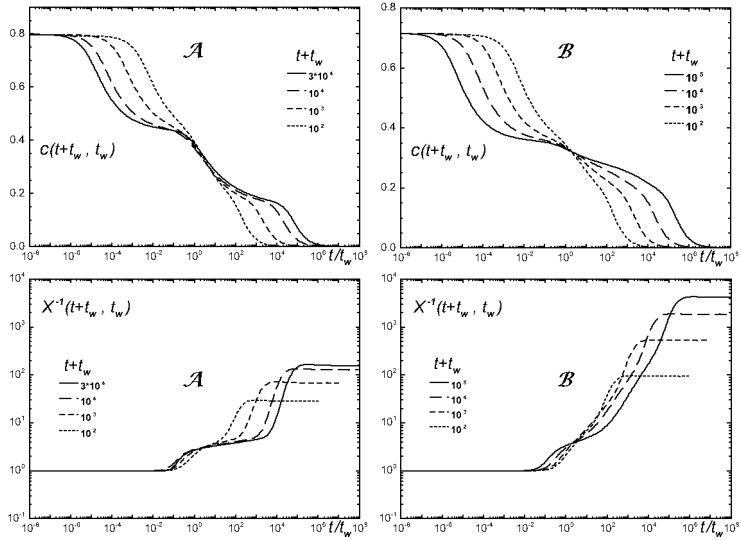} 
\caption{Correlation function $c(t+t_w,t_w)=q(t+t_w,t_w)-m(t+t_w)\,m(t_w)$ and FDT-violation
parameter $X(t+t_w,t_w)$ \ for $T=0.5$, \ $h=0.5$ (${\cal A}$-phase) and 
$h=0.7$ (${\cal B}$-phase) \cite{Ho03} corresponding to point a and b in Fig.\ref{PhaseDiax}. 
The curves are plotted for fixed $t+t_w$. The range $t/t_w\sim t+t_w$ corresponds to
$t_w\sim 1$.}
\label{cX(t,tw)-AB} 
\end{figure}

In the ${\cal A}$-phase, for $t\gg\tau_p$ and $t_w\gg 1$, the correlation function approaches a 
scaling form
\begin{equation}
c(t+t_w,t_w)\to \hat c_a(t/\tau_a) \qquad {\rm with}\quad \tau_a\approx t_w\,,
\end{equation}
whereas no such scaling is found in the ${\cal B}$-phase. This is in some
analogy to one step versus full \index{replica symmetry breaking}replica symmetry breaking
\cite{Parisi-rev}.

Within replica theory the
$p\!\!=\!\!3$-spin-glass\index{p-spin-glass} does, however, not exhibit a full replica symmetry
broken phase at all. This is another difference in the phase diagrams obtained from dynamics
and replica theory. For the same range of time, the FDT-violation
parameter
$X(t+t_w,t_w)$ develops a plateau in the ${\cal A}$-phase, consistent with the
QFDT-solution\index{QFDT}, whereas no such plateau shows up in the ${\cal B}$-phase.

\subsection{Mean Field Dynamics of Systems with Short Ranged Interactions}
\index{short ranged interactions}\label{ntaw}

The slowing down of the dynamics near $T_c$ in a disordered system is not expected to be
associated with a diverging length scale, at least not on the level considered here. A diverging
length scale is expected only for certain four point correlation functions,
which are factorized in the dynamic mean field theory or in mode coupling theory, e.g.
\begin{equation}
C(R_{ij},t\!-\!t')=\ol{\av{\phi_i(t)\phi_j(t)\phi_i(t')\phi_j(t')}}^J 
\!-\!\ol{\av{\phi_i(t)\phi_i(t')}\av{\phi_j(t)\phi_j(t')}}^J .
\end{equation}

Dynamic mean field equations for short ranged interactions can be derived adopting a
factorization property
\begin{equation}\label{nfao}
q_{i,j}(t,t')=f_{i,j}\cdot q(t,t') \qquad\quad r_{i,j}(t,t')=f_{i,j}\cdot r(t,t').
\end{equation}
The corresponding factorization has been proposed and tested within mode coupling theory for
supercooled liquids \cite{MCT,Sch03}. The resulting equations for the time dependent parts,
$q(t,t')$ and $r(t,t')$, are of the same form as those derived for mean field models,
Eqs.(\ref{hfre}-\ref{nfeu}). The definition of $W(q)$, \req{jrso}, has to be modified, taking into
account the short range nature of the interactions and the form factors $f_{i,j}$. The crucial
assumption is, to neglect four point correlations of the form given above.

It should be mentioned, that a rather different picture, based on a \index{droplet model}droplet
model description, has been developed \cite{FH86}. Both pictures, mean field models and the
droplet model, reproduce certain features of short ranged spin-glasses, and no clear cut evidence
in favour or against one of the theories seems to exist at present.

\subsection{$p$-Spin-Glass with Relaxing Bonds and Cage Relaxation in Supercooled Liquids}
\index{cage relaxation} 

In a \index{supercooled liquid}supercooled liquid one may distinguish two kinds of motion.  Each
particle in a sense is surrounded by a cage formed by other particles. Within this cage it may
vibrate with amplitudes much smaller than the interparticle spacing. This picture is of course
simplified in the sense, that the cage is also vibrating, and each particle is also part of the
cage of other particles. It is probably more appropriate to view this motion as a collective
localized anharmonic vibration. The second type of motion is a rearrangement of the particles
forming the cage, or an escape of a particle from its cage. This type of motion involves jumps
over distances of the order of the interparticle spacing, and this is supposed to be an activated
process \cite{Dont,Stil82}.

The mode coupling theory for supercooled glasses \cite{MCT,Sch03} deals with density
fluctuations, and the resulting equations are identical to those derived for the
$p\!=\!3$-spin-glass \index{p-spin-glass} in equilibrium, \req{gtem}. The initial decay of the
correlation function towards the plateau (see Fig.\ref{q(t)>Tc}) is usually interpreted as being
due to the motion within the cages, whereas the ultimate decay is supposed to result from the
escape of a particle from its cage. On the other hand, activated processes are not contained in
the standard mode coupling theory. The extended theory \cite{FGo92} captures those processes, and
yields a rounding of the singularities found in the ideal mode coupling theory at $T_c$. 
This extended theory is still an equilibrium theory and can not describe aging \cite{Ree91}.

In the following I sketch a slightly different picture \cite{Ho03} resulting in identical mode
coupling type of equations above $T_c$. This formulation allows, however, to deal with off
equilibrium properties. It is based on a distinction of the two types of motion described above. 

We neglect for a moment the activated process of rearrangement of the cages and consider the
anharmonic vibrations only. The position of a particle at time $t$ may be written as 
\begin{equation}
\vec{x}_i(t)=\vec{R}_i+\vec{u}_i(t) 
\end{equation}
where $\vec{u}_i(t)$ is the momentary displacement of particle $i$ from its mean position
$\vec{R}_i$. The configuration of the $\vec{R}_i$ can be viewed as inherent state \cite{Stil82}. 
The dynamics of the displacements $\vec{u}_i(t)$
 is treated within the selfconsistent anharmonic phonon
theory \cite{Ho67}, developed originally for quantum crystals or strongly anharmonic
\index{selfconsistent phonon theory} crystals at high temperatures. The main idea of this
approach is, to replace the harmonic and anharmonic coupling constants by their averages over
thermal or quantum fluctuations, e.g.
\begin{equation}
\av{V}_{ij}=\int\!\d r\,V''(r)\,g_{ij}(r).
\end{equation}
$V(r)$ is the pair potential and $g_{ij}(r)$ the static pair correlation function
between particles $i$ and $j$. Employing the \index{factorization property}factorization property
\req{nfao} and retaining cubic anharmonicities, one ends up with equations identical to those of
the dynamic mean field theory, Eqs.(\ref{hfre}-\ref{nfeu}). 

The coupling constants $\av{V}_{i\cdots}$ are calculated for fixed mean positions $\vec{R}_i$.
These positions are, however, random and the coupling constants are therefore random variables as
well. They play the role of the $J_{i\cdots}$ in \req{oygn}. Taking into account the activated
motion of the $\vec{R}_i(t)$, the random couplings $J_{i\cdots}(t)$ now depend on time as well.
Under the assumption that they are Gaussian distributed random variables, \req{btes} holds in
the modified form
\begin{equation}
\ol{J_{i\cdots}(t) J_{j\cdots}(t')}
=\sfrac{1}{p!N^{p/2}}\{\delta_{i,j}\cdots+\cdots\}W_p G_J(t-t').
\end{equation}
Since the couplings $J_{\cdots}(t)$ are dynamical variables, corresponding correlation functions
have to be introduced
\begin{equation}
\ol{J_{i\cdots}(t) \hat J_{j\cdots}(t')}
=\sfrac{1}{p!N^{p/2}}\{\delta_{i,j}\cdots+\cdots\}W_p F_J(t-t').
\end{equation}
The resulting mean field equations are unchanged, except for the modified memory terms,
\req{bgsx},
\begin{eqnarray}
K(t,t')&=&W''\big(q(t,t')\big)\,G_J(t-t')\,r(t,t')+W'\,\big(q(t,t')\big)\,F_J(t-t')
\nonumber\\
M(t,t')&=&W'\,\big(q(t,t')\big)\,G_J(t-t').
\end{eqnarray}
Assuming an activated process for the cage relaxation,
\begin{equation}
G_J(t)=\e^{-t/\tau_{\rm cage}} \qquad {\rm and}\qquad 
F_J(t)=\frac{\beta_J}{\tau_{\rm cage}}\e^{-t/\tau_{\rm cage}}
\end{equation}
is chosen. 

The choice $\beta_J=0$ and $\tau_{\rm cage}\to\infty$ is a realisation of quenched
disorder \cite{Ho84}. The results for $t_w\gg\tau_{\rm cage}$ are similar to those obtained for
aging in Sect.\ref{ojrb}, with $\tau_{\rm cage}$ replacing $t_w$. This holds in particular for the
algebraic decay of the correlation function  $q(t)-q_c\sim t^{-a}$.

The situation for $\beta_J=\beta$ is quite different. This means that the dynamics of the cages
(bonds) can also equilibrate, and that fluctuation-dissipation-theorems hold for
$t_w\gg\tau_{\rm cage}$. Numerical integration of the dynamic mean field equations yields results
shown in  Fig.\ref{q(t,tcage)}.\\
\begin{figure}[ht]
\centering
\includegraphics[width=8cm]{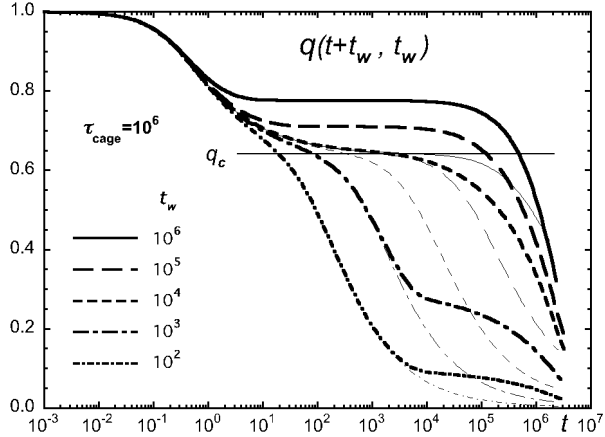} 
\caption{Correlation-function $q(t+t_w,t_w)$ for $T<T_c$, cage relaxation time $\tau_{\rm
cage}=10^6$  and various waiting times. For comparison the correlation function for quenched
disorder ($\tau_{\rm cage}=\infty$) is also shown (thin lines).}
\label{q(t,tcage)} 
\end{figure}\\
There are now two long time scales, the waiting time $t_w$ and the cage
relaxation time $\tau_{\rm cage}$. There is yet another characteristic time scale
$\hat\tau_{\rm cage}=\tau_p(\tau_{\rm cage})$ associated with the cage relaxation time, where
$\tau_p$ is the plateau time introduced in Sect.\ref{ojrb}.

For $t+t_w<\hat\tau_{\rm cage}$ the results found for quenched
disorder are recovered. This means that the approach and departure of the correlation-function
from the plateau $q_c$ is ruled by power laws, fluctuation-dissipation-theorems are violated for
$t>\tau_p(t_w)$ and aging is observed. 

For $t_w\gg\tau_{\rm cage}$ the system equilibrates, a new plateau value
$>q_c$ is found. The ultimate decay of the correlation function is now ruled by $\tau_{\rm
cage}$. No critical behaviour, i.e. no power laws in $t$, are found. The exponential decay of the
correlation function towards the new plateau value is associated with a time scale 
$\tau_p(T)\sim (T_c-T)^{-1}$.
\begin{figure}[ht]
\centering
\includegraphics[width=8cm]{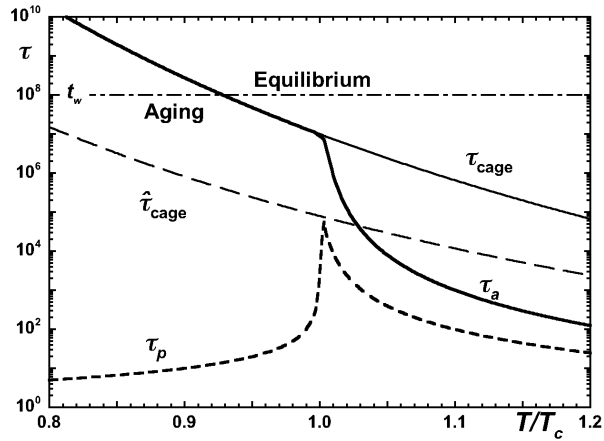} 
\caption{Temperature dependent time scales for the $p$-spin-glass \index{p-spin-glass} with
relaxing bonds. For the waiting time $t_w$ indicated in the figure the equilibrium and aging
regimes are marked}
\label{t_cage} 
\end{figure}

The \index{susceptibility}susceptibility $\chi''(\omega)$, calculated for the $p$-spin-glass model
with relaxing bonds, depends on the assumptions made for $F_J(t)$ or $\beta_J$. For equilibrated
cage relaxation
$\beta_J=\beta$ a ``knee'' shows up at $\omega_p\sim 1/\tau_p$, and $\chi''(\omega)\sim \omega$
for $\omega<\omega_p$ whereas $\chi''(\omega)\sim \omega^a$ for $\omega>\omega_p$. Such a knee has
actually been observed in CKN \cite{CKN92} but this observation was later withdrawn. The knee is
absent assuming non equilibrated cage dynamics with $F_J=0$.
\begin{figure}[ht]
\centering
\includegraphics[width=8cm]{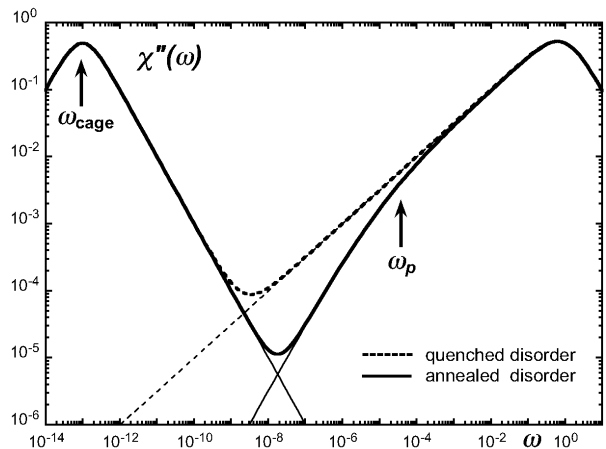} 
\caption{Susceptibility $\chi''(\omega)$ for the $p$-spin-glass with relaxing bonds. ``Quenched
disorder'' calculated with $\beta_J=0$, ``annealed disorder'' calculated with $\beta_J=\beta$.}
\label{X(omega)-J} 
\end{figure}

It has to be mentioned that the relaxation of the cage configurations has been put in by hand. In
particular it is not influenced by the critical slowing down of the anharmonic vibrations. This
is a consequence of the mean field model and the $N$-dependent scaling of the interactions, 
\req{htev}. More realistic models with short range interactions would yield a coupling of the 
dynamics of the interactions $J$ and the anharmonic vibrations, even within the mean field
approximation discussed in Sect.\ref{ntaw}. This has, however, not been worked out.

\begin{figure}[ht]
\centering
\includegraphics[width=3.5cm]{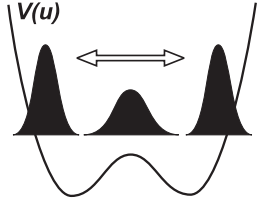} 
\caption{Coupled 1-2-phonon oscillation in an anharmonic potential.}
\label{1-2-phonon} 
\end{figure}

The picture outlined in this section attributes the dynamics of a supercooled liquid, on a time
scale shorter than the life-time of cages, to anharmonic vibrations. The instability at $T_c$ is
attributed to a softening of a coupled oscillation of the center of mass and the width of the
thermal cloud, representing the motion of a particle in its cage. This motion is sketched in
Fig.\ref{1-2-phonon}. The idea of a \index{coupled 1-2-phonon process}coupled 1-2-phonon process
was originally developed for quantum crystals \cite{Ho67}. The critical slowing down near $T_c$
within this picture is not directly related to the glass transition, and the actual glass
transition at $T_g$ has to be attributed to a slowing down of the activated cage relaxation.

The standard interpretation \cite{Kir87,Sch03,Cug03} is, to identifying the dynamic trans\-ition
temperature $T_c$ with the crossover region in glasses, and the onset temperature of replica
symmetry breaking $T_{\rm RSB}$ with the Kauzmann temperature, which has to be lower than the
glass transition temperature $T_g$.  A comparison of $T_c$ and $T_g$ for various glass
formers, as done in Fig.\ref{Angell}, reveals that both temperatures can be rather far apart,
whereas model calculations \cite{Kra99} result in differences between $T_c$ and $T_{RSB}$ much
smaller than the values observed.

\section{Outlook}
In this lecture I have concentrated on the dynamics of disordered systems treated in mean field
theory. I have discussed models within physics but also models for various questions outside
physics. Among those are optimization problems, neural networks and problems related to economics.
Especially for the applications outside physics, where spacial dimension is not relevant, mean
field theory is in many cases exact. The dynamic mean field theory is quite rich because the order
parameters are functions of two time variables, the correlation- and response-functions. At high
temperature or strong noise level, the systems are ergodic. Lowering the temperature or noise
level, a freezing transition shows up and the system becomes non ergodic.

Even on the level of mean field dynamics there are open questions, for instance the arbitrariness
in the time parametrization function $h(t)$ discussed in Sect.\ref{ojrb}, or the scaling form of
correlation- and response-functions in the aging regime of the ${\cal B}$-phase, discussed in
Sect.\ref{vbrt}. A more thorough understanding of the relationship between dynamics and replica
theory is desired. In particular there is no obvious prescription of how to calculate entropy or
free energy from dynamics. These quantities are, however, of prime interest in replica theory.

The coarse features of aging in spin-glasses are reasonably well described by dynamic mean field
theory. There are, however, more elaborate experiments investigating special cooling and heating
schedules \cite{Vin99}. It is not clear whether they can be understood on the basis of dynamic
mean field theory.

Glasses and spin-glasses are systems with short ranged interactions. Some of their properties are
captured by dynamic mean field theory or replica theory. Spacial correlations are not taken into
account and questions concerning the existence of diverging length scales can not be answered.
On the other hand real space formulations, e.g. the droplet model for spin-glasses \cite{FH86} or
models with kinetically constraint dynamics for glasses \cite{Jaec91,Ber03}, are concerned
with spacial aspects and a better understanding of discrepancies and common features would be
desired.

After all, it is surprising that the dynamic mean field theory for disordered systems can be
applied to so many systems in physics and in other disciplines.\\[4mm]
{\footnotesize{\bf Acknowledgment}: Partially supported by the ESF-programme SPHINX.}



\begin{theindex}

  \item A-phase, 23
  \item aging, 1, 4, 7, 13, 21

  \indexspace

  \item B-phase, 23, 24

  \indexspace

  \item cage relaxation, 26
  \item combinatorial optimisation, 1, 9
  \item continuous transition, 18
  \item coupled 1-2-phonon process, 30
  \item creep, 1, 8
  \item critical exponent, 20, 23
  \item crossover region, 2
  \item crossover scaling, 19, 23

  \indexspace

  \item discontinuous transition, 18, 23
  \item drift of a particle, 8, 23, 24
  \item droplet model, 26
  \item dynamic mean field theory, 16

  \indexspace

  \item econophysics, 1, 10
  \item effective temperature, 22
  \item ergodic components, 2
  \item ergodic phase, 17

  \indexspace

  \item factorization property, 27
  \item FDT-violation, 22, 24
  \item field cooled susceptibility, 3
  \item fluctuation-dissipation-theorem, 14, 17
  \item freezing temperature, 2
  \item freezing transition, 13
  \item frustration, 3

  \indexspace

  \item glass transition, 5
  \item glassy dynamics, 1
  \item Glauber dynamics, 5, 11
  \item graph bipartitioning, 9

  \indexspace

  \item Ising model, 4

  \indexspace

  \item K-sat problem, 10

  \indexspace

  \item Langevin dynamics, 13

  \indexspace

  \item marginal stability, 23
  \item mean field model, 11
  \item memory terms, 17
  \item minority game, 10
  \item mode coupling theory, 2, 6, 21

  \indexspace

  \item neural network, 1, 8, 23

  \indexspace

  \item p-spin-glass, 2, 6, 7, 11, 25, 26, 29
  \item path integral representation, 13
  \item phase diagram, 18, 23
  \item pinning, 1, 8
  \item plateau, 19, 23

  \indexspace

  \item QFDT, 23, 25
  \item QFDT-phase, 21, 23
  \item quenched disorder, 15

  \indexspace

  \item reparametisation invariance, 24
  \item replica symmetry breaking, 1, 12, 25
  \item replica theory, 1
  \item replica trick, 12

  \indexspace

  \item selfconsistent phonon theory, 27
  \item short ranged interactions, 25
  \item simulated annealing, 1, 9, 10
  \item SK-model, 11
  \item spherical model, 11
  \item spin-glass, 1, 2
  \item stability criterium, 18
  \item supercooled liquid, 1, 5, 26
  \item susceptibility, 6, 20, 21, 28

  \indexspace

  \item ultrametricity, 2

  \indexspace

  \item waiting time, 1

  \indexspace

  \item zero field cooled susceptibility, 3

\end{theindex}

\end{document}